\documentclass[%
 reprint,
nofootinbib,
 amsmath,amssymb,
 aps,
floatfix,
]{revtex4-1}

\usepackage{graphicx}% Include figure files
\usepackage{dcolumn}% Align table columns on decimal point
\usepackage{bm}% bold math
\usepackage[colorlinks=true]{hyperref}% add hypertext capabilities
\usepackage[normalem]{ulem}

\begin{document}

\preprint{APS/123-QED}

\title{Probing hybrid stars with gravitational waves via interfacial modes}

\author{Shu Yan Lau} \email{sl8ny@virginia.edu}
\author{Kent Yagi} \email{ky5t@virginia.edu}

\affiliation{Department of Physics, University of Virginia, Charlottesville, Virginia 22904, USA}

\date{\today}

\begin{abstract}
One of the uncertainties in nuclear physics is whether a phase transition between hadronic nuclear matter to quark matter exists in supranuclear matter equations of state. Such a feature can be probed via gravitational-wave signals from binary neutron star inspirals that contain information of the induced tides. The dynamical part of the tides is caused by the resonance of pulsation modes of stars, which causes a shift in the gravitational-wave phase. In this paper, we investigate the dynamical tides of the interfacial mode ($i$-mode) of spherical degree $l=2$, a non-radial mode caused by an interface associated with a quark-hadron phase transition inside a hybrid star. In particular, we focus on hybrid stars with a crystalline quark matter core and a fluid hadronic envelope. We find that the resonant frequency of such $i$-modes typically ranges from 300Hz to 1500Hz, and the frequency increases as the shear modulus of the quark core increases. We next estimate the detectability of such a mode with existing and future gravitational-wave events from the inspiral waveform with a Fisher analysis. We find that GW170817 and GW190425 have the potential to detect the $i$-mode if the quark-hadron phase transition occurs at sufficiently low pressure and the shear modulus of the quark matter phase is large enough. We also find that the third-generation gravitational-wave detectors can further probe the $i$-mode with intermediate transition pressure. This finding opens a new, interesting direction for probing the existence of quark core inside a neutron star.

\end{abstract}

\maketitle

\section{Introduction}

The equation of state (EOS) of matter in the high-density, low-temperature regime remains uncertain until now. Although quantum chromodynamics (QCD) allows us to theoretically predict the properties of matter, it can be solved perturbatively only at asymptotic densities~\cite{Kurkela:2009gj}. At densities below the nuclear saturation density, chiral effective field theory (CFT) is useful for obtaining the EOSs to a good precision (\cite{1990PhLB..251..288W,1991NuPhB.363....3W,BOGNER200559}, see \cite{RevModPhys.81.1773} for a review). However, at intermediate densities, around 1-10 times the nuclear saturation density, the non-perturbative nature of QCD makes it difficult to constrain the microscopic theory of matter, while CFT calculations fail to converge at these energy scales. Meanwhile, neutron star (NS) cores, where densities lie within this range, serve as natural laboratories that allow us to probe the properties of cold dense matter.

QCD predicts that matter undergoes a phase transition from hadronic matter to quark matter at high density. The possibility of the existence of deconfined quark matter inside NS cores has been of interest for decades. Recently, Annala et~al. \cite{Annala:2019puf} showed some evidence supporting deconfinement within massive NSs. This particular class of NSs containing quark matter cores, known as hybrid stars (HSs)~\cite{Alford_2005}, may show unique signatures in their observables due to the existence of a phase transition between hadronic and quark matter. For instance, a strong phase transition can lead to the existence of twin stars (\cite{PhysRev.172.1325,SCHERTLER2000463,Paschalidis:2017qmb}), i.e. a HS and a NS with the same mass but different radii. 

It is expected that within a low-temperature, high-density environment like the HS core, the quarks form Cooper pairs and exist in a color superconducting phase due to the attractive channels of the strong interaction. At asymptotically high densities where the up, down, and strange quarks have negligible mass, it is well-established that the nine quarks of different flavors and colors pair up equally to form Cooper pairs through the BCS mechanism and exist in the color-flavor-locked (CFL) phase \cite{ALFORD1999443}. For the relatively lower density region where the strange quark mass becomes more significant, other color superconducting phases are proposed. One possibility is the crystalline color superconducting (CCS) phase \cite{PhysRevD.65.094022,PhysRevD.63.074016}. This phase is formed from the pairing of quarks with unequal magnitudes of momenta through a non-BCS mechanism, causing the spontaneous breaking of translational invariance. Its crystalline properties are studied in \cite{PhysRevD.76.074026}, which predicts that its shear modulus can be as high as 1000 times that of a NS crust\footnote{The astrophysical properties of systems having such a rigid phase are investigated in  \cite{PhysRevLett.99.231101,PhysRevD.76.081502,PhysRevD.77.023004,PhysRevD.79.083007,PhysRevD.88.124002,PhysRevD.89.103014,PhysRevD.95.101302,Pereira:2020cmv}.}. 

The EOS has been constrained by measuring the macroscopic NS parameters (e.g., mass, radius). The majority of the observed pulsars through electromagnetic (EM) signals have masses between 1--2~$M_\odot$~\cite{Ozel:2016oaf}, with the most massive one measured to be $2.14^{+0.10}_{-0.09}$~$M_\odot$ \cite{2020NatAs...4...72C}. The mass and radius measurement from thermonuclear bursts and quiescent low-mass x-ray binaries have been used to probe the EOS~\cite{Steiner:2010fz,Lattimer:2013hma,Lattimer:2014sga,Ozel:2015fia,Steiner:2017vmg}. Recently, Neutron star Interior Composition Explorer (NICER) measured the mass and radius of PSR J0030+0451~\cite{Miller:2019cac,Riley:2019yda}, which has also been used to constrain the EOS further~\cite{Raaijmakers:2019qny}.

The features in the EOS can also be inferred from gravitational-wave (GW) observations through tides. 
During the late inspiral stage, the equilibrium part of tides leaves an imprint on the GW phase characterized by the tidal deformability \cite{Flanagan:2007ix,Vines:2011ud,Damour:2012yf}, which is the quasi-static linear response coefficient. Such tidal deformability can be used to probe HSs~\cite{Paschalidis:2017qmb,Nandi:2017rhy,Zhou:2017pha,Montana:2018bkb,Burgio:2018yix,Annala:2019puf,Chatziioannou:2019yko,Miao:2020yjk,Parisi:2020qfs}. Close to merger, dynamical part of tides becomes important~\cite{Hinderer:2016eia}.
One can also use post-merger signals to probe HSs~\cite{Most:2018eaw,Bauswein:2018bma,Bauswein:2020aag}.

The GW events GW170817 \cite{PhysRevLett.119.161101} and GW190425 \cite{Abbott_2020} have placed constraints on the weighted-averaged tidal deformability parameter, $\bar{\Lambda}$. Moreover, the normalized tidal deformability of a 1.4~$M_\odot$ NS is found to have an upper bound of 800 within a 90~\% confidence level for the low spin scenario in GW170817  \cite{PhysRevLett.119.161101}. One can further constrain the tidal deformability \cite{Chatziioannou:2018vzf,Abbott:2018exr} by using theoretical knowledge of certain quasi-universal relations \cite{I-Love-Q-Science,I-Love-Q-PRD,Yagi:2015pkc,Yagi:2016qmr,Yagi:2016bkt}. These tidal deformability measurements of NSs can be mapped to bounds on the NS radius~\cite{Annala:2017llu,Bauswein:2017vtn,De:2018uhw,Most:2018hfd} and constrain the EOS (see e.g. \cite{PhysRevLett.119.161101,Abbott:2018wiz,Zhang:2018bwq,Tews:2018chv,Landry:2018prl,Essick:2019ldf,Carson:2018xri,Raithel:2019ejc}). Moreover, one can combine various multi-messenger observations of NSs. For example, Dietrich \textit{et~al.} \cite{2020arXiv200211355D} combined  the GW signal from GW170817, the NICER observation of PSR J0030+0451 and radio observations of PSR J0740+6620, PSR J0348+4032, PSR J1614-2230 to find a new constraint on the radius of a 1.4~$M_\odot$ NS as $11.74^{+0.98}_{-0.79}$~km within 90~\% confidence level (see also e.g. \cite{Radice:2017lry,Radice:2018ozg,Nandi:2018ami,Kiuchi:2019lls,Raaijmakers_2020,Zimmerman:2020eho,Essick:2020flb} for other constraints on the NS tidal deformability, radius and EOS with multi-messenger observations).

The dynamical tides are the resonance of the quasi-normal modes of the NSs, which can cause a phase shift in the GW signal as the orbital frequency sweeps through the resonant frequency of each mode \cite{10.1093/mnras/270.3.611}. 
The dominant non-radial pulsation mode is the fundamental mode ($f$-mode). This mode has a relatively high resonant frequency of $>1500$~Hz and is excited at a very late stage of inspiral or in the post-merger phase in which the two NSs merge and form a massive NS. The frequency is beyond the sensitive region of the current ground-based detectors, making its detection from the GW signal very challenging. 

On the other hand, there are modes with lower resonant frequencies, such as the gravity modes ($g$-modes), which fall within the most sensitive part of the detectors' spectral noise curve. These modes depend on the internal properties like composition gradient or temperature (see, e.g.,  \cite{1992ApJ...395..240R,1994ApJ...426..688R,1980tsp..book.....C,Kokkotas_1999,Wei:2018tts}). The discontinuity $g$-mode is a special type of $g$-mode caused by the existence of a discontinuity in density, which can occur at the interface between the hadronic and quark phase inside a HS~\cite{10.1093/mnras/227.2.265,Tonetto:2020bie}\footnote{An analogy to this mode is the deep water gravity waves at the water-air interface on Earth.}. This mode is sometimes referred to as the interfacial mode ($i$-mode) in certain literature studying NSs with a crust \cite{1988ApJ...325..725M,PhysRevD.92.063009,2020arXiv201209637P} and it can be caused by discontinuities in either the density or the shear modulus. In particular, McDermott \textit{et~al.} \cite{1988ApJ...325..725M} studied NSs with a solid crust and surface ocean and showed that there is one $i$-mode associated with each of the core-crust interface and crust-ocean interface at a fixed spherical harmonic order $l$. 
In this paper, we use the terminology ``$i$-mode" instead of ``discontinuity $g$-mode" for the mode associated with the quark-hadron interface inside a HS to avoid confusion with other $g$-modes associated with factors like composition gradient. 

Recent studies have shown the potential detectability of $g$-modes in various NS models with different compositions using third-generation GW detectors by stacking multiple events \cite{10.1093/mnras/stx1188}. It is known that the properties of the non-radial mode spectrum in HSs with a first-order phase transition can be quite different from those of NSs \cite{PhysRevD.65.024010,Ranea_Sandoval_2018,Orsaria_2019}. Whether these modes in a HS are detectable by the current detectors is certainly of interest. Since the $i$-mode (or the discontinuity $g$-mode) depends strongly on the properties of the quark-hadronic matter interface, detection of such a mode would provide strong evidence of deconfinement within the HSs. The goal of this paper is to study the detectability of the $i$-mode of various HS EOSs with a first-order phase transition from the GW signal of a HS-HS merger, assuming the quark matter core is in the CCS phase.

In this paper, we first calculate the $i$-modes of a set of HS models with a first-order phase transition between the hadronic phase and the CCS quark phase. We then analyze the detectability of the $i$-mode resonance from an inspiralling HS binary using Fisher analysis. 
We assume the HS core to be in the CCS phase to incorporate the effect of shear modulus on the $i$-mode to investigate not only the effect of density discontinuity, but also that of the solid-fluid transition on the detectability with GW observations.

We find that the detectability depends on the transition pressure ($P_t$) of the EOSs. For low $P_t$ models, the $i$-mode of some models is detectable with Advanced LIGO (aLIGO) at its design sensitivity. As $P_t$ increases, the effect of $i$-mode on the GW phase decreases and becomes less detectable even with the next-generation detectors. Besides, we consider the effect of the shear modulus of the CCS quark matter core on the $i$-mode detectability. We find that the $i$-mode resonant frequency and phase shift magnitude both increase with the shear modulus. While the increased phase shift makes the mode easier to detect for models with low $P_t$ EOSs, the increased resonant frequency reduces its detectability for models with higher $P_t$ if the frequency exceeds the inspiral cutoff frequency. 
Using gravitational waveform parameters corresponding to GW170817 and GW190425, we also find that the $i$-mode can potentially be detected for low $P_t$ EOSs if the quark matter shear modulus is high enough. Among the two events, the $i$-mode is less detectable with GW190425 than GW170817 due to its larger luminosity distance.

This paper is organized as follows: In Sec.~\ref{Sec:mode_cal}, we describe the method to calculate the background HS models with the general relativistic equations and the non-radial pulsation modes within Newtonian theory. We then consider the dynamical tides within the inspiralling HSs which leads to the excitation of the $i$-modes. In Sec.~\ref{Sec:Fisher}, we describe the method for parameter estimation with the Fisher information matrix, which allows us to quantify the detectability of the $i$-mode parameters. In Sec.~\ref{Sec_EOS}, we describe the HS EOSs to be considered in this study. In Sec.~\ref{sec:results}, we calculate the detectability of $i$-modes with the method described in the previous sections. In Sec.~\ref{Sec:ConsistencyCheck}, we check the consistency of our method, which combines the relativistic calculation for the background model with the Newtonian calculation of the pulsation modes and the tidal coupling, with another approach purely within the Newtonian formalism.
We conclude in Sec.~\ref{sec:conclusion} and present possible future directions.

\section{\label{Sec:mode_cal} Mode contribution to waveforms}

In this section, we first explain how to compute the $i$-mode oscillations of HSs via a hybrid method. We next describe how such oscillation modes affect the GW waveforms from binary HS inspiral.

\subsection{Non-radial pulsation modes in a hybrid formalism}

To calculate the effects of the $i$-modes on the GW signal, we need to first solve for the $i$-mode frequencies and eigenfunctions for a given EOS and use this to find the tidal coupling coefficient which will be introduced in Sec~\ref{tidal_coupling}. The formulation within the Newtonian framework is described in \cite{10.1093/mnras/270.3.611}, which requires one to use the Newtonian equations to construct the background solution and the perturbed, pulsating solution. In this paper, we take a different approach called a hybrid formulation (see, e.g.,  \cite{PhysRevLett.108.011102,10.1093/mnras/stx1188}), where we include fully relativistic effects for the background but keep the perturbation to a Newtonian level.

For the background non-rotating, radially-symmetric  solution, we solve the Tolman-Oppenheimer-Volkoff (TOV) equations given by\footnote{Note that we do not need to solve for the $(t,t)$ component of the background metric since we apply the Newtonian pulsation equations and the unperturbed Newtonian potential is simply given by $-m/r$.} 
\begin{align}
    \frac{d P(r)}{d r} &= -\frac{\left(\rho + P\right) \left(m + 4\pi r^3 P\right)}{r^2 \left(1 - 2m / r \right)}, \\
    \frac{d m(r)}{d r} &= 4 \pi \rho r^2,
\end{align}
where $m(r)$ is the mass enclosed within a sphere of radius $r$ from the stellar center. Integrating the above equations together with the EOS and requiring that the pressure vanishes at the stellar surface, we obtain the static profile of the HS.

The formulae governing the pulsation in Newtonian theory can be found in various literature. We employ the formulation in \cite{1988ApJ...325..725M}, without taking the Cowling approximation, i.e. without omitting the gravitational perturbations. The formulation and the corresponding derivation are briefly discussed in Appendix~\ref{Newt_puls}. By numerically solving the set of pulsation equations, we can obtain the eigenfrequencies and eigenfunctions of a set of non-radial modes for each spherical degree $l$.

\subsection{Tidal coupling and phase shift in the waveform} \label{tidal_coupling}

During a HS-HS inspiral, the $i$-modes resonates as the orbital frequency sweeps through the resonant frequency and causes a phase shift in the GW waveform. Following Lai \cite{10.1093/mnras/270.3.611}, the overall phase shift for an $l=2$ mode is given by the equation
\begin{align}
    \delta \phi_\alpha = -\frac{5 \pi^2 }{4096} \left(\frac{R}{M}\right)^5 \frac{2q}{\left(1+q\right) } \frac{1}{\Omega_{n2m}^2}\left| Q_{n2m} \right|^2,
\end{align}
where $M$ and $R$ are the stellar mass and radius, $q$ is the ratio of the companion mass to that of the pulsating HS, $\Omega_{n2m}$ is the normalized resonant frequency for the $l=2$ mode defined by
\begin{align}
    \Omega_{nlm}^2 = \frac{R^3\omega_{nlm}^2}{M},
\end{align}
with $\omega_{nlm}$ representing the mode angular frequency. $Q_{nlm}$ is the tidal coupling coefficient defined by
\begin{align}
    Q_\alpha = Q_{nlm} = \frac{1}{MR^l} \int d^3 x \rho \, \vec{\xi}^*_{nlm} \cdot \nabla \left( r^l Y_{lm} \right). \label{eq:Tidal_Coef}
\end{align}
Here we use the set of subscripts $\alpha = ( n, l, m)$ to specify an eigenmode with a radial quantum number $n$, spherical harmonics degree $l$ and order $m$. The quantum number $n$ is an index that labels all the non-radial modes with the same $l$ and $m$, ranked in ascending order of resonant frequencies. For a typical NS with a solid crust, this includes the fundamental ($f$) mode, the interfacial ($i$) mode and the gravity ($g_1$, $g_2$, ...) modes, etc.\footnote{$f$ and $i$ do not have any subscripts since for each ($l$,$m$) there is only one $f$-mode and one $i$-mode per interface.}~\cite{1988ApJ...325..725M}. The eigenvectors are normalized by 
\begin{align}
    \int d^3 x \rho \left|\vec{\xi}_{nlm}\right|^2 = MR^2. \label{eq:eigenmode_norm}
\end{align}

We investigate only the $\{l,m\}$ = $\{2,\pm2\}$ $i$-mode contribution on the GW phase, which dominates the phase shift. From Eq.~\eqref{eq:Tidal_Coef}, we can easily see that $Q_{n22}=Q_{n2-2}$. Hence, we have the $i$-mode overall phase shift given by
\begin{align}
    \delta \phi_{\alpha_i} = -\frac{5 \pi^2 }{2048} \left(\frac{R}{M}\right)^5 \frac{2q}{\left(1+q\right) }  \frac{1}{\Omega_{n_i22}^2} \left| Q_{n_i22} \right|^2, \label{eq:phase_from_Q}
\end{align}
where $n_i$ is the radial quantum number corresponding to the $i$-mode, and $\alpha_i$ is the index representing the combined contributions from the $l=2$ $i$-modes, i.e., the sum of $\{n_i,2,2\}$ and $\{n_i,2,-2\}$ modes. 

The contribution from a pulsation mode on a binary inspiral waveform appears as a shift in the phase and time when the binary sweeps through the resonant frequency. The resulting correction to the phase in the frequency domain is given by \cite{PhysRevD.75.044001,10.1093/mnras/stw2552,10.1093/mnras/stx1188}
\begin{align}
    \Delta\Psi_{\alpha_i}(f) = - \sum_{A=1,2}\delta\phi_{\alpha_i}^{(A)} \left( 1 - \frac{f}{f_{\alpha_i}^{(A)}}\right) \theta(f - f_{\alpha_i}^{(A)}), \label{phi_resonance}
\end{align}
where $\Delta\Psi_{\alpha_i}(f)$ is the phase correction in frequency domain, $\delta\phi_{\alpha_i}^{(A)}$ and $f_{\alpha_i}^{(A)}$ are the overall phase shift and the resonant frequency due to the $i$-mode of the $A$th body, and $f$ is the GW frequency from the inspiral. $\theta(f - f_{\alpha_i})$ is the Heaviside step function. To reduce the number of parameters, we follow~\cite{Pan:2020tht} and rewrite the above phase shift as
\begin{equation}
\Delta\Psi_{\alpha_i}(f) \approx - \delta \bar \phi_{\alpha_i} \left( 1 - \frac{f}{\bar f_{\alpha_i}}\right) \theta(f - \bar f_{\alpha_i}),
\label{phi_resonance2}
\end{equation}
where the total phase shift $\delta\bar{\phi}_{\alpha_i}$ and the weight-averaged mode frequency $\bar{f}_{\alpha_i}$ are given by
\begin{align}
    \delta\bar{\phi}_{\alpha_i}  &= \delta\phi^{(1)}_{\alpha_i}+\delta\phi^{(2)}_{\alpha_i}, \label{eq_phia}\\
    \bar{f}_{\alpha_i} &= \delta\bar{\phi}_{\alpha_i} \left( \frac{\delta\phi^{(1)}_{\alpha_i}}{f^{(1)}_{\alpha_i}}+\frac{\delta\phi^{(2)}_{\alpha_i}}{f^{(2)}_{\alpha_i}}\right)^{-1}. \label{eq_fa}
\end{align}
In the following, we drop the subscript $\alpha_i$ on the mode frequency and phase shift to simplify the expressions.

\section{\label{Sec:Fisher}Fisher analysis}

For a signal with a high signal-to-noise ratio (SNR), we use the Fisher information matrix to approximate the posterior distribution of the GW signal parameters~\cite{Cutler:1994ys}. Given a GW signal $h(t)$ that depends on a set of parameters contained in the vector $\theta^a$, the Fisher matrix is defined by 
\begin{align}
    \Gamma_{ab} = \left( \frac{\partial h}{\partial \theta^a}\bigg|\frac{\partial h}{\partial \theta^b}\right),
\end{align}
where the inner product between $a(t)$ and $b(t)$ is defined by
\begin{align}
    \left(a\big|b\right) = 2\int_{0}^{\infty} \frac{\Tilde{a}^* \Tilde{b}+\Tilde{a} \Tilde{b}^*}{S_n(f)} df.
\end{align}
Here, the overhead tilde represents the Fourier transform while $*$ represents a complex conjugate. $S_n(f)$ is the spectral noise density of the detector. For simplicity, we follow~\cite{Cutler:1994ys,Berti:2004bd} and assume the prior of $\theta^a$ to be a Gaussian function with a root-mean-square $\sigma_a$. The effective Fisher matrix taking into account this information is given by 
\begin{align}
    \Tilde{\Gamma}_{ab} = \left( \frac{\partial h}{\partial \theta^a}\bigg|\frac{\partial h}{\partial \theta^b}\right) + \frac{1}{\sigma_a^2}\delta_{ab}.
\end{align}
The root-mean-square uncertainty in the measurement of $\theta^a$ is given by 
\begin{align}
    \Delta \theta_a = \sqrt{\Sigma_{aa}}\,, \quad \Sigma_{ab} \equiv  \left(\Tilde{\Gamma}^{-1}\right)_{ab}\,.
\label{Fisher_uncertainty}
\end{align}
If the uncertainty is smaller than the measured value in magnitude, it is considered detectable. It is also convenient to define the correlation coefficients to quantify the correlations between different parameters:
\begin{align}
 C_{ab}=\frac{\Sigma_{ab}}{\sqrt{\Sigma_{aa} \Sigma_{bb}}}. \label{Eq:corr}
\end{align}
The diagonal element of $C_{ab}$ is normalized to unity while the off-diagonal elements quantify the amount of correlation between two different parameters, ranging from 0 (no correlation) to $\pm 1$ (strong correlation).

The frequency domain waveform has the form
\begin{align}
    h(f)= A(f) e^{-i\Psi(f)}.
\end{align}
 The functional forms of the amplitude $A(f)$ and phase $\Psi(f)$ depend on the waveform templates. In this paper, we use the sky-averaged “IMRPhenomD” GW waveform template \cite{PhysRevD.93.044006,PhysRevD.93.044007} for point particles, with the addition of the 5PN and 6PN tidal contributions to the phase in~\cite{Vines:2011ud,PhysRevD.89.103012}, as well as the effect of mode resonance given in Eq.~\eqref{phi_resonance2}. The elements of the parameter set $\theta^a$ are given by
\begin{align}
    \theta^a = \left( \ln{\mathcal{A}}, \phi_c, t_c, \ln{\mathcal{M}_z}, \ln{\eta}, \chi_s, \chi_a, \bar{\Lambda}, \delta \bar{\Lambda}, \bar f, \delta\bar \phi \right).
\end{align}
The meaning of each element is as follows: the sky-averaged normalized amplitude 
\begin{align}
\mathcal{A}= \frac{\mathcal{M}_z^{5/6}}{\sqrt{30}\pi^{2/3} D_L};
\end{align}with the luminosity distance from the source $D_L$; the redshifted chirp mass $\mathcal{M}_z = \mathcal{M} \left(1+z\right)$, where 
\begin{equation}
\mathcal{M}= \frac{(m_1 m_2)^{3/5}}{(m_1+m_2)^{1/5}}
\end{equation}
is the chirp mass; the symmetric mass ratio 
\begin{equation}
\eta = \frac{m_1 m_2}{(m_1+m_2)^2};
\end{equation}
the symmetric and asymmetric spin parameters $\chi_{s,a}=\left(\chi_1 \pm \chi_2\right)/2$, where $\chi_{1,2}$ are the dimensionless spins of the individual stars; the reparametrization of the mass weighted tidal deformabilities (see e.g.~\cite{PhysRevD.89.103012})
\begin{align}
\bar{\Lambda} =& \frac{8}{13}\Big[\left(1+7\eta-31\eta^2\right)\left(\Lambda_1+\Lambda_2\right) \nonumber\\
&+\sqrt{1-4\eta}\left(1+9\eta-11\eta^2\right)\left(\Lambda_1-\Lambda_2\right)\Big],\\
\delta\bar{\Lambda} =& \frac{1}{2}\Big[\sqrt{1-4\eta}\left(1 - \frac{13272}{1319}\eta+\frac{8944}{1319}\eta^2\right)\left(\Lambda_1+\Lambda_2\right),\nonumber\\
&+\left(1 - \frac{15910}{1319}\eta+\frac{32850}{1319}\eta^2+\frac{3380}{1319}\eta^3\right)\left(\Lambda_1-\Lambda_2\right)\Big],
\end{align}
where $\Lambda_{1,2}$ are the individual tidal deformabilities normalized by $m_{1,2}^5$; 
the phase-shift-weighted $i$-mode frequency $\bar f$ and the overall phase shift due to the $i$-mode $\delta \bar \phi$. Note that if we consider binaries of identical stars, we  have $\bar{\Lambda}=\Lambda_1=\Lambda_2$ and $\delta\bar{\Lambda}=0$. 

At high frequencies, the tidal part of the waveform that we use becomes less accurate as the HSs will eventually come to contact. Following \cite{Cutler:1994ys}, we only consider the inspiral waveform, which terminates at a separation of $6(m_1+m_2)$, which is equivalent to the radius of the innermost stable circular orbit (ISCO) of an object orbiting around a non-spinning central object with mass $(m_1+m_2)$. This corresponds to a cutoff frequency $f_{\text{ISCO}} = \left[6^{3/2} \pi (m_1+m_2)\right]^{-1}$ in the Fisher estimate\footnote{For stiff EOSs, the HSs may come to contact before reaching the separation of $6(m_1+m_2)$, i.e., $R_1+R_2 > 6(m_1+m_2)$, where $R_1$ and $R_2$ are the radii of the HSs. In these cases, the actual cutoff frequency should be set lower than $f_{\text{ISCO}}$. However, since the spectral noise density increases quickly in the high-density region, the uncertainty estimates using the Fisher matrix is not significantly affected as long as the actual cutoff frequency does not differ too much from $f_{\text{ISCO}}$ and the $i$-mode resonant frequencies are not too close to the cutoff frequency.}. 

In the following analysis, we pick the fiducial values for ($\phi_c$, $t_c$, $\chi_s$, $\chi_a$) to be (0, 0, 0, 0). The tidal deformability parameters ($\bar{\Lambda}$, $\delta\bar{\Lambda}$) are set as (800, 0) for identical HS binaries \footnote{In reality, $\bar{\Lambda}$ varies for different EOSs. However, we have checked that $\Delta \delta \bar \phi$ is insensitive to the choice of $\bar{\Lambda}$  and thus in this study, we fix its value to be 800 for simplicity . Same applies to $\delta \bar{\Lambda}$ with its value fixed to be 0.} and are specified otherwise in asymmetric cases. We use the spin priors of $\left|\chi_{s,a}\right| < 1$ and tidal priors of $0 < \bar{\Lambda} < 3000$ and $|\delta \bar{\Lambda}| < 500$ \cite{PhysRevD.89.103012}. The values of $\bar{f}$ and $\delta \bar{\phi}$ depend on the HS models and are calculated with the method described in Sec.~\ref{Sec:mode_cal}.

\section{\label{Sec_EOS} Equation of state}

\begin{figure*}[!htb]
    \centering
    \begin{minipage}{.5\textwidth}
        \centering
        \includegraphics[width=8.6cm]{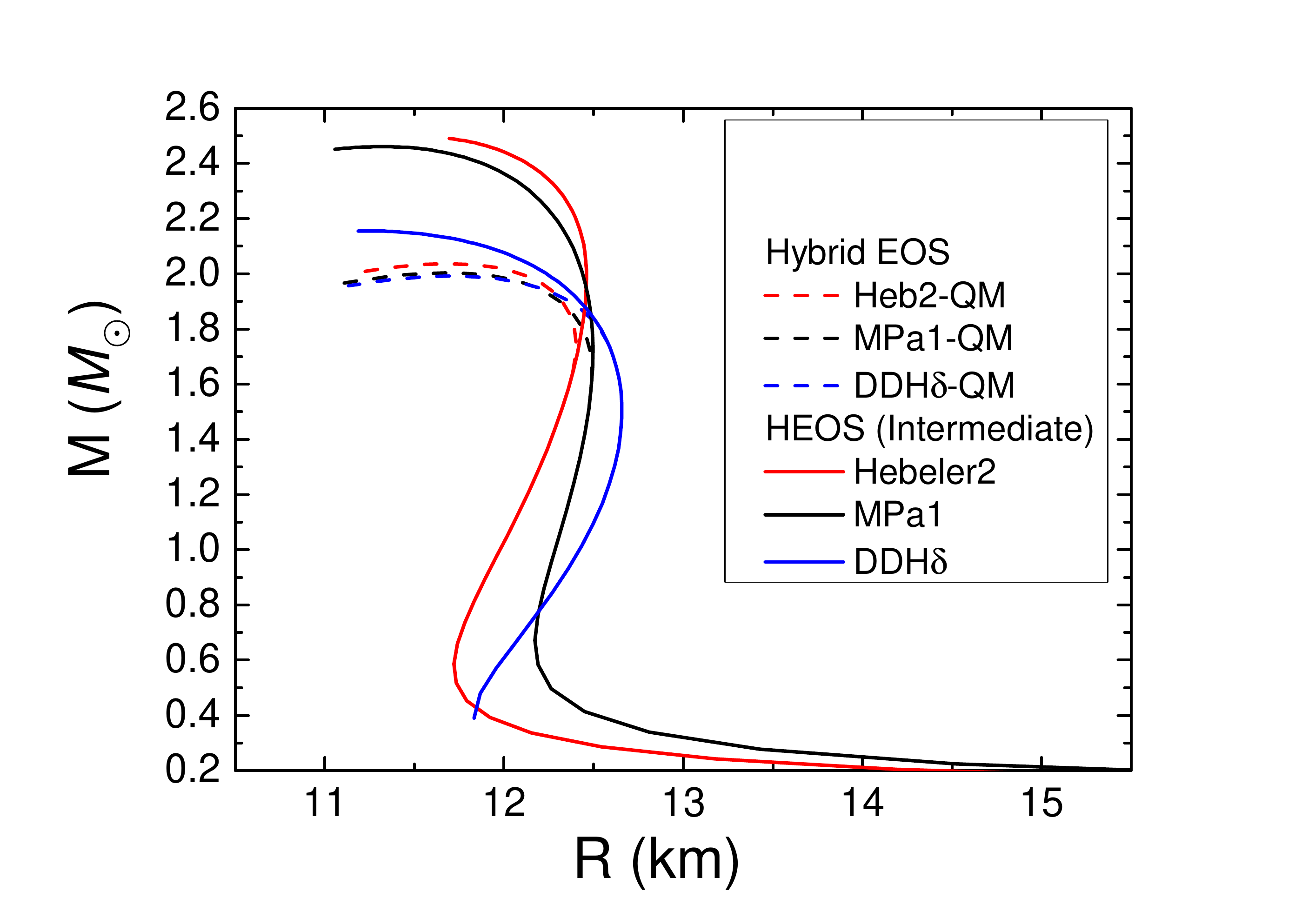}
    \end{minipage}%
    \begin{minipage}{.5\textwidth}
        \centering
        \includegraphics[width=8.6cm]{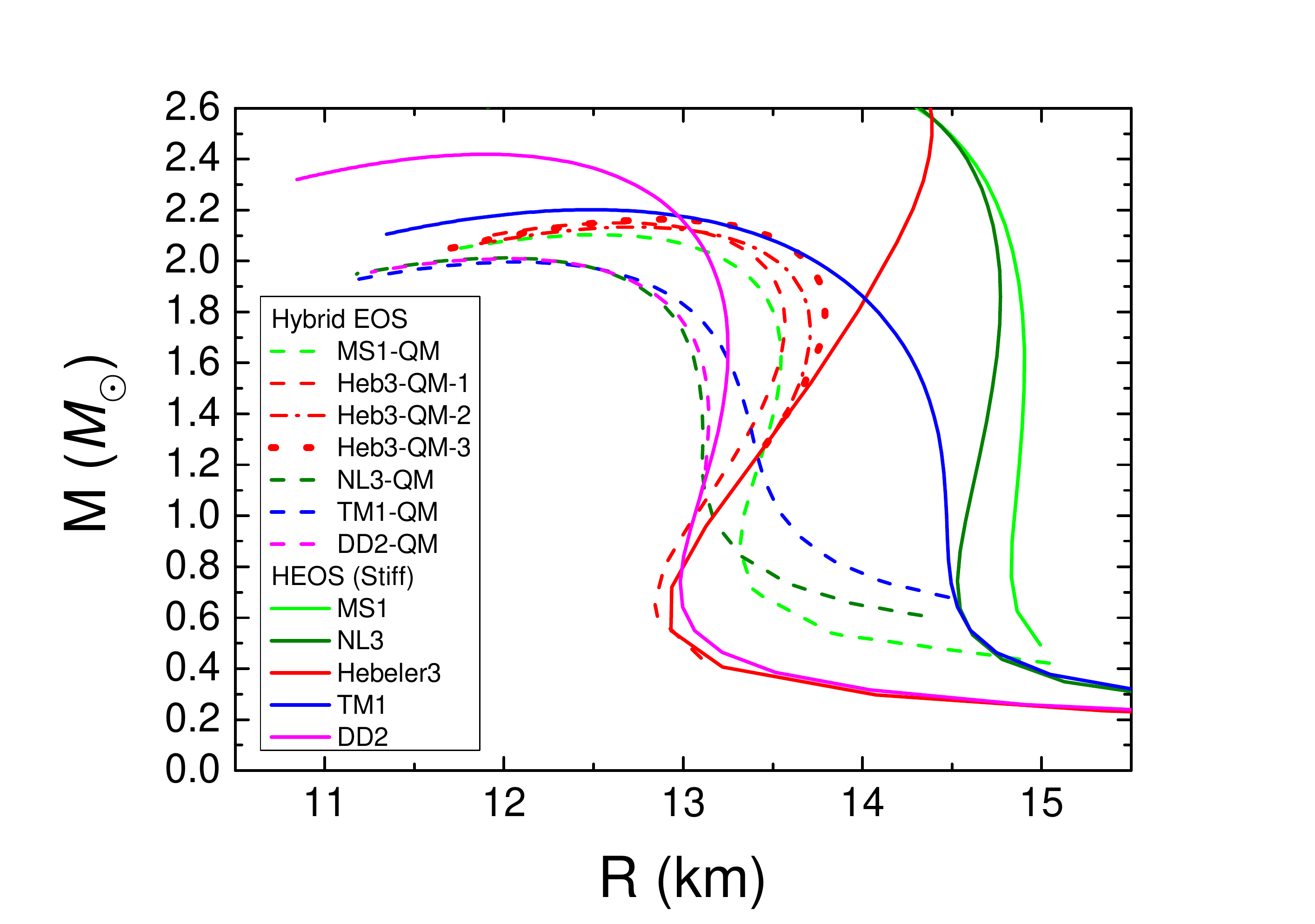}
    \end{minipage}%
    \caption{$M$-$R$ relations of the HS models (dashed lines) and hadronic matter models (solid lines) constructed with intermediate (left) and stiff (right) HEOSs.}
    \label{Figure_1}
\end{figure*}

Let us now describe how we construct the EOSs used in our analysis.

\subsection{\label{ssec_qm_eos} Quark matter EOS: Modified Bag Model}

The quark matter EOS is described by the Bag model \cite{Alford_2005}:
\begin{align}
\Omega = -P = -\frac{3}{4\pi^2} a_4 \mu_q^4 + \frac{3}{4\pi^2} a_2 \mu_q^2 + B_{\text{eff}}, \label{Equation1}
\end{align}
where $\Omega$ is the grand potential density, $P$ is the pressure, $\mu_q$ is the quark chemical potential, and $(a_4, a_2, B_{\text{eff}})$ are phenomenological parameters. The physical meaning and the ranges of the parameters are discussed in \cite{Alford_2005}. The parameter $a_4$ accounts for the QCD coupling constant and takes a value between 0 and 1. $a_2$ is the contributions from both the pairing gap of the color-superconducting phase and the strange quark mass. $B_{\text{eff}}$ is the effective bag constant that models confinement. The value of $a_2$ is expected to be of order $10^4\left(\text{MeV}\right)^2$. In the case of the simplest MIT Bag model consisting only of free massless quarks, the bag constant, $B_{\text{MIT}}$, lies within the range of $ 145~\text{MeV} < B_{\text{MIT}}^{1/4} < 160~\text{MeV}$ (see, e.g., \cite{2007ASSL..326.....H} and references therein). In the modified Bag model, it is instead treated as an arbitrary parameter. 

The other thermodynamic variables is determined using Eq.~(\ref{Equation1}) and thermodynamic relations. In particular, the energy density $\rho$ is given by
\begin{align}
    \rho = \frac{9}{4\pi^2} a_4 \mu_q^4 - \frac{3}{4\pi^2} a_2 \mu_q^2 + B_{\text{eff}}. 
\end{align}
We assume the quark matter core to be in the CCS phase, a non-BCS color superconducting phase existing as an extremely rigid solid. The shear modulus is given by the formula \cite{PhysRevD.76.074026}
\begin{align}
    \mu = \nu_0 \left( \frac{\Delta}{10~\text{MeV}} \right)^2 \left( \frac{\mu_q}{400~\text{MeV}} \right)^2, \label{CCSC_mu}
\end{align}
where the constant $\nu_0$ has a value of $2.47~\text{MeV}/\text{fm}^{3}$. $\Delta$ is the gap parameter of the CCS phase with a range between 5~MeV and 25~MeV \cite{PhysRevD.76.074026}.

\subsection{\label{ssec:HEOS}Hadronic matter EOS}

Next, we explain hadronic EOSs (HEOSs) for constructing HSs. It is expected that the EOS gets softened as quark matter appears inside the core. To ensure that the HS EOSs have the maximum stable mass beyond the 2~$M_\odot$ constraint from observations, we do not consider HEOSs that are too soft. The models we classify as intermediate in terms of stiffness are: MPa1 \cite{MUTHER1987469}, DDH$\delta$ \cite{GAITANOS200424}, Hebeler2; and those with high stiffness are: MS1 \cite{MULLER1996508}, NL3 \cite{PhysRevC.55.540}, TM1 \cite{SUGAHARA1994557}, Hebeler3. The models Hebeler2 and Hebeler3 are taken from the subtables labelled as ``intermediate" and ``stiff" respectively in Table~5 of \cite{Hebeler_2013}. They are the representative HEOSs with the low-density part satisfying the results derived from CFT. The sub- and supranuclear density parts satisfy the constraints from massive pulsars. For simplicity, we do not include detailed crust models containing additional phase transitions and possible density gaps in the outer crust region that can give rise to additional $i$-modes or $g$-modes in the low-frequency region (10-100~Hz) \cite{PhysRevD.92.063009}.

\subsection{Hybrid star models}

We now use the quark and hadronic matter EOSs explained in the previous subsections to construct HS models.
The first-order phase transition from hadronic matter to quark matter is modeled with Maxwell construction, which requires the continuity of pressure and the baryon chemical potential, assuming local charge neutrality. The density is discontinuous at the transition point. The procedure of the construction is presented in Appendix~\ref{Maxwell_Construction}.

We construct HS models with different combinations of $a_4$, $a_2$, $B_{\text{eff}}$, $\Delta$ and nuclear matter EOSs, requiring the HS EOSs to satisfy the observational constraints on the maximum mass ($M_\mathrm{TOV}>2~M_\odot$), radius ($R_{1.4 M_\odot}\in[8.9,13.5]$~km 
from various multimessenger observations; see Table 1 of \cite{2020arXiv200211355D}) and tidal deformability ($\bar{\Lambda}<800$). 
The EOS parameters of the HS models constructed are listed in Table~\ref{Table_1}. 
\begin{table}[!ht]
\begin{ruledtabular}
\begin{tabular}{cccccc}
EOS    & $a_4$	& $a_2^{1/2}$ & $B_{\text{eff}}^{1/4}$ & HEOS & $P_t$ \\ 
 & & $(\text{MeV})$ & $(\text{MeV})$ &  & $(\text{dyn cm}^{-2})$ \\\hline
MS1-QM  & 0.52 & 108 & 135 & MS1   & 6.21311E33 \\
Heb3-QM-1  & 0.5  & 102 & 134 & Hebeler3  & 1.14143E34 \\
NL3-QM & 0.53 & 90  & 140 & NL3 & 1.03622E34 \\
TM1-QM & 0.55 & 105 & 140 & TM1 & 1.32036E34\\ \hline
Heb3-QM-2 & 0.53 & 143 & 128 & Hebeler3  & 5.01506E34 \\
Heb3-QM-3  & 0.53 & 156 & 123 & Hebeler3  & 4.75498E34 \\
DD2-QM & 0.55 & 100 & 140 & DD2 & 4.32026E34 \\\hline
MPa1-QM  & 0.57 & 90 & 140  & MPa1 & 1.23181E35 \\
Heb2-QM & 0.55 & 70 & 140  & Hebeler2      & 1.25885E35 \\
DDH$\delta$-QM  & 0.57 & 87 & 142  & DDH$\delta$  & 1.53379E35 
\end{tabular}
\end{ruledtabular}
\caption{\label{Table_1} 
HS EOSs with the quark matter EOS parameters, the HEOSs and $P_t$ for the envelope listed. The EOSs are divided into 3 sections characterized by the transition pressures: low $P_t$ (top), intermediate $P_t$ (middle), high $P_t$ (bottom).}
\end{table}

In Fig.~\ref{Figure_1}, we show the mass-radius relations of the HS models and the HEOSs. We classify the EOSs into ``intermediate" and ``stiff" EOSs based on their radius within the mass range between 1--2 $M_\odot$. We do not consider HEOSs that are too soft, since the appearance of quark matter softens the EOSs further for densities beyond the quark-hadron transition point compared to the corresponding HEOSs, which leads to a maximum stable mass below the current bound of 2~$M_\odot$. The quark matter EOS parameters are also restricted within a certain range due to this maximum mass constraint.

\begin{figure*}[t]
    \centering
    \includegraphics[width=8.6cm]{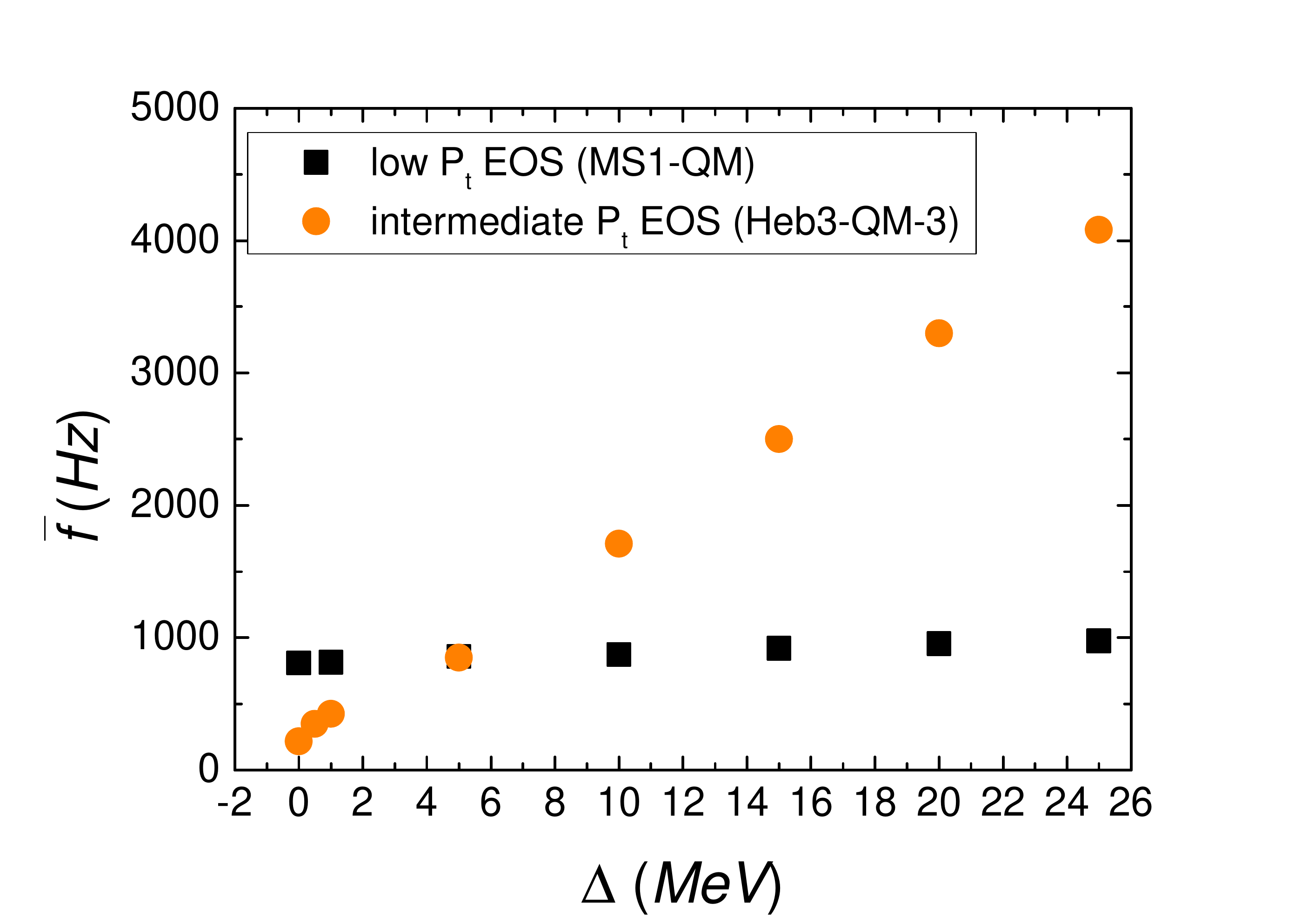}
        \includegraphics[width=8.6cm]{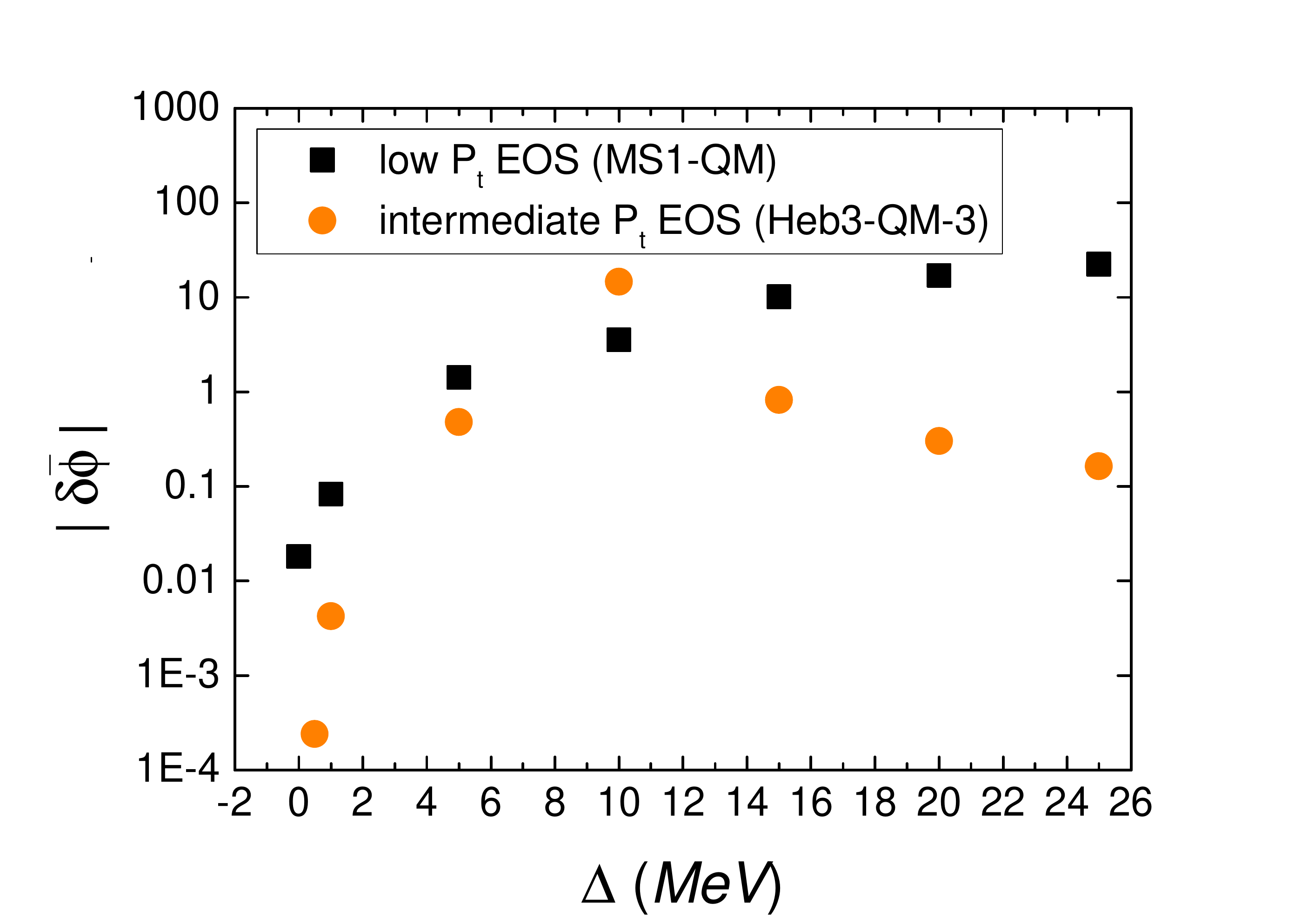}
    \caption{(Left) The weight-averaged $i$-mode frequency of (1.4,1.4)~$M_\odot$ HS binary models against $\Delta$. A low $P_t$ EOS (MS1-QM; in black squares) and an intermediate $P_t$ EOS (Heb3-QM-3; in orange dots) are chosen to construct the models.
    (Right) Similar to the left panel but for the total overall phase shift. The phase shift of Heb3-QM-3 near $\Delta = 10$~MeV exceeds over 10 due to the avoided crossing between the $i$-mode and the $f$-mode (not shown in this figure). Near this region, the phase shift of the two modes comes close to each other and the resonant frequencies repel to avoid a degeneracy.
    }
    \label{Figure_f_vs_gap}
\end{figure*}

Putting the observational constraints into consideration, we expect that the transition pressure $P_t$ of the HS EOSs in Table~\ref{Table_1} is loosely correlated with the stiffness of the HEOSs in order to produce models that are stiff enough to support 2~$M_{\odot}$, but cannot be too stiff not to exceed the upper bound set on the tidal deformability and radius measurements. Roughly speaking, the maximum mass observations constrain the EOS stiffness from below while the upper bound set on radius and tidal deformability measurements constrain from above. Since the appearance of quark matter softens the EOSs and generally lowers the maximum mass, soft HEOSs that barely meet the constraints on maximum mass cannot be used to construct valid HS EOSs. Also, an HS EOS with a hadronic part of intermediate stiffness must have a high $P_t$, which in turn gives a transition point at relatively high mass on the $M-R$ curve (see the left panel of Fig.~\ref{Figure_1}), to satisfy the 2~$M_\odot$ lower bound on the maximum mass. Meanwhile, those with a stiff HEOS cannot have a high $P_t$, or else it would exceed the 13.5~km upper bound on the radius (see the right panel of Fig.~\ref{Figure_1}). 

In the following analysis, we classify the HS EOSs according to $P_t$. The models with HEOSs of intermediate stiffness will have a high $P_t$ in order to meet the observational constraints. For those with a stiff HEOS, we can construct a wider range of $P_t$ covering intermediate $P_t$ and low $P_t$ as indicated in Table~\ref{Table_1}, while the HS models still satisfy the observation bounds.

\section{\label{sec:results}Results}

Let us now present all the numerical results. We first show how the $i$-mode frequency and phase shift depend on the quark parameters, in particular $\Delta$. We next present the detectability of such modes with current and future GW observations, including the existing GW events of GW170817 and GW190425.

\subsection{$i$-mode dependence on the properties of the phase transition}

The frequency and phase shift of the $i$-mode depend strongly on both the density gap and shear modulus gap at the interface. Each EOS listed in Table~\ref{Table_1} has a specific value of density gap, while the shear modulus for each model can still vary with $\Delta$ according to Eq.~(\ref{CCSC_mu}). To get an idea of how the elastic properties affect the $i$-mode, we consider HSs with quark matter in the CCS phase with different $\Delta$s. 

In Fig.~\ref{Figure_f_vs_gap}, we show the $i$-mode frequency and phase shift against $\Delta$ of two representative HS models with 1.4~$M_\odot$. MS1-QM, denoted by the black squares, is a HS model with a low $P_t$, while Heb3-QM-3, denoted by orange dots, is a model with intermediate $P_t$. 

The $\Delta$ dependence of $\bar{f}$ and $\delta \bar{\phi}$ are found to be much stronger for the intermediate $P_t$ model. Besides, the $\Delta$ dependence for $\delta \bar{\phi}$ for this model is not monotonic in contrast to the low $P_t$ models. This is because $\delta \bar{\phi}$ varies as the square of the tidal coupling coefficient, $Q_{nlm}$, and inversely with the square of the mode frequency (see Eq.~\eqref{eq:phase_from_Q}). For small $\Delta$, the rate of increase in $|Q_{nlm}|$ outweighs that of the mode frequency, while the opposite happens at large $\Delta$.  This causes $\delta \bar{\phi}$ to increase initially and fall off for large $\Delta$.

\begin{figure*}%[!hpbt]
    \includegraphics[width=8.6cm]{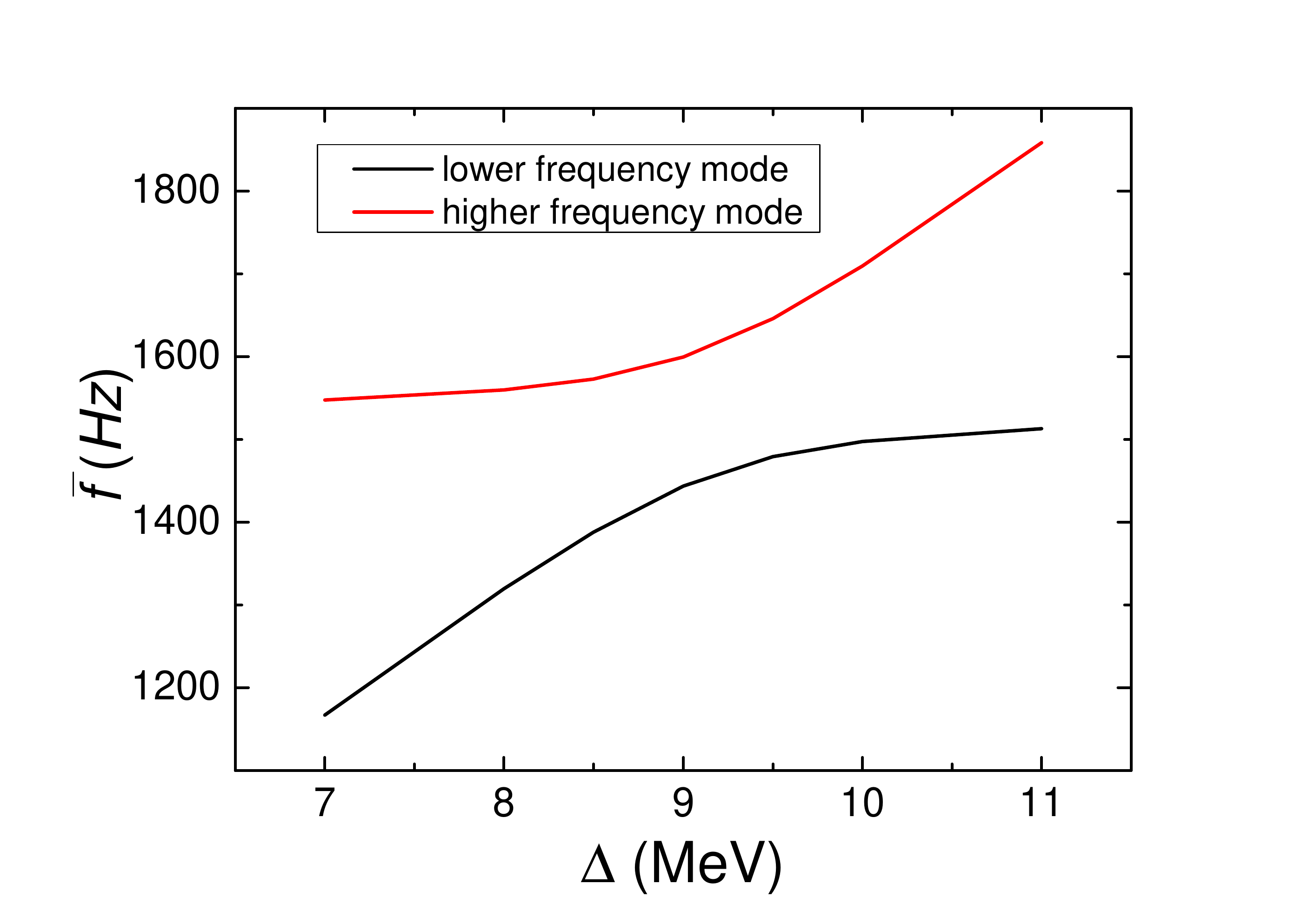}
    \includegraphics[width=8.6cm]{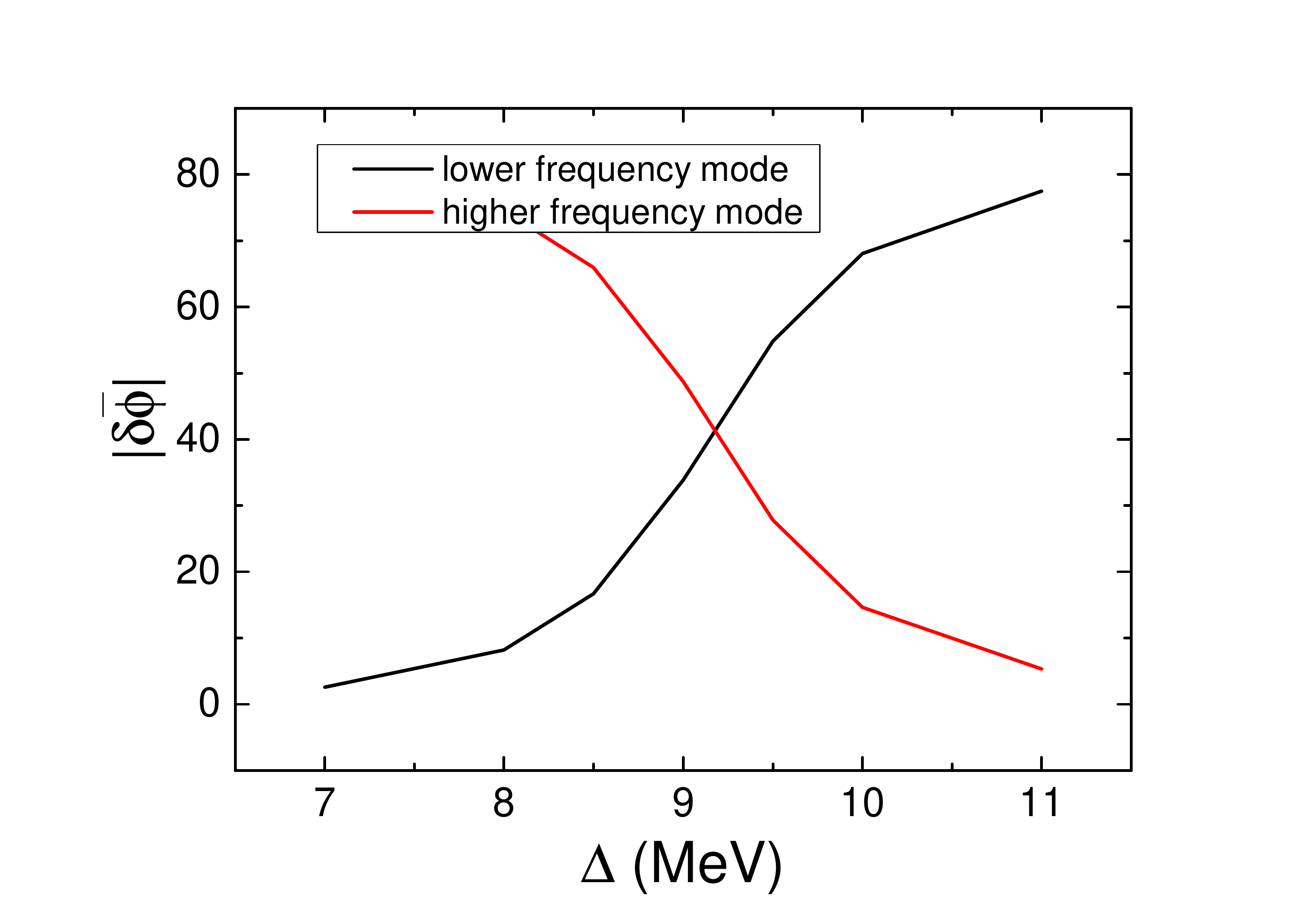}
 \caption{\label{avoided_crossing} The avoided crossing of the $i$-mode and $f$-mode as $\Delta$ changes from 7 to 11~MeV of a 1.4~$M_\odot$ HS model with the EOS Heb3-QM-3. The left panel shows the repulsion of the frequencies of the higher frequency mode and the lower frequency mode. The right panel shows the exchange in $|\delta \phi|$ between the two modes.}
 \end{figure*}

The peak of $|\delta \bar{\phi}|$ with a value of $\sim 40$ near $\Delta$ = 10~MeV is a consequence of mode repulsion. When the frequency of the $i$-mode is close to another mode, such as a spheroidal shear mode, the mode frequencies repel with each other without crossing while the phase shift of the two modes comes close to each other. This phenomenon is the avoided crossing and is commonly observed in stellar pulsation problems (see e.g., Ch.17 of \cite{1980tsp..book.....C}) as well as other eigenvalue problems. The avoided crossing near $\Delta=10$~MeV in Fig.~\ref{Figure_f_vs_gap} happens between the $i$-mode and the $f$-mode. 
To further demonstrate this phenomenon, we show in Fig.~\ref{avoided_crossing}  both the $i$-mode and $f$-mode for $\Delta$ between 7 and 11~MeV with the EOS Heb3-QM-3. Observe that there is a repulsion in mode frequencies around $\Delta = 9.5$~MeV, while $|\delta \bar{\phi}|$ of the two modes cross each other.

Before we discuss the detectability of the $i$-mode of HSs, it is worth pointing out its difference from the typical $i$-modes associated with the interface(s) inside a NS (such as the one between the hadronic fluid envelope and the solid crust or phase transitions inside the crust). The $i$-mode frequency of HSs typically ranges between 300 and 1500~Hz, while that of a NS is generally lower. Kr\"uger \textit{et~al.} \cite{PhysRevD.92.063009} computed the $l=2$ $i$-modes of a NS with the SLy4 EOS and a crust model with multiple first-order phase transitions using a general relativistic formalism. All of the $i$-modes have frequencies below 121~Hz. From the difference in mode frequencies, the $i$-mode of a HS can be clearly distinguished from that of a NS.

\subsection{$i$-mode detectability with gravitational waves}

Upon the observation of a GW signal, one can estimate the parameters that ``best fit" the waveform to the measured signal buried inside the noise. Due to this, the estimated parameters always come with uncertainties. In Sec.~\ref{Sec:Fisher}, we have briefly discussed how the parameter estimation errors can be found using the Fisher matrix for a large SNR. In particular, the statistical uncertainties of $\bar{f}$ and $\delta \bar{\phi}$ in the parameter estimation determine whether they are measurable from the signal. As our numerical result shows that the relative uncertainty in $\delta \bar{\phi}$ is always larger than that of $\bar{f}$, the detection criterion of the $i$-mode can therefore be set as $\Delta(\delta \bar{\phi}) < \left|\delta \bar{\phi}\right|$.

\subsubsection{Equal-mass systems}

Let us first analyze the detectability of the phase shift due to the excitation of the $i$-mode during the inspiral of a \emph{symmetric} (equal-mass, non-spinning) HS-HS merger, which consists of identical HSs. With this assumption, $\bar{f}$ is identical to the $i$-mode resonant frequency, and $\delta \bar{\phi}$ is twice the phase shift of the individual HS. If we fix the mass of our models and assume no spin, there are 4 parameters that depend on the EOSs: the tidal deformability parameters $\bar{\Lambda}$ and $\delta \bar{\Lambda}$, the (weighted-averaged) $i$-mode resonant frequency $\bar{f}$, and the overall orbital phase shift $\delta \bar{\phi}$. One might expect the detectability of the $i$-mode to depend on all of the parameters. However, we found that the correlation between the tidal deformability parameters and the $i$-mode parameters is small. For example, the correlation coefficient
%$C_{ab}$ using Eq.~\eqref{Eq:corr}. 
$C_{\bar{\Lambda} \, \delta\bar{\phi}}$ defined in Eq.~\eqref{Eq:corr} is about 0.001--0.03 and similar for other combinations between the tidal deformability and $i$-mode parameters, which is much lower than that for the correlation between $\bar f$ and $\delta \bar \phi$ $(C_{\bar{f} \, \delta\bar{\phi}} \sim 0.3-0.8)$. Hence, we can estimate the detectability by varying fiducial values of $\bar{f}$ and $\delta \bar{\phi}$ only, keeping those of $\bar \Lambda$ and $\delta \bar \Lambda$ fixed.

\begin{figure*}
    \centering
    \includegraphics[width=8.6cm]{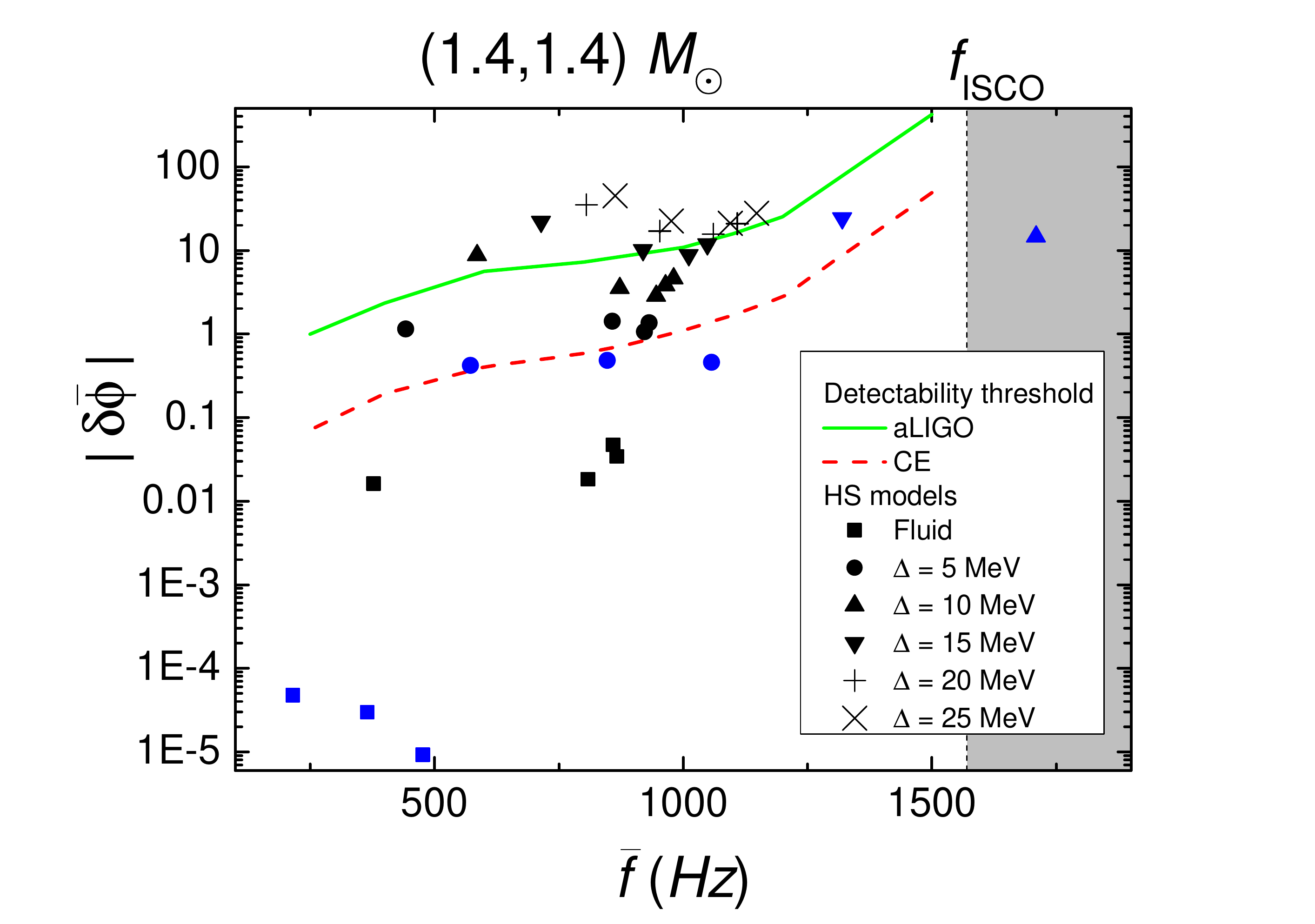}
 \includegraphics[width=8.6cm]{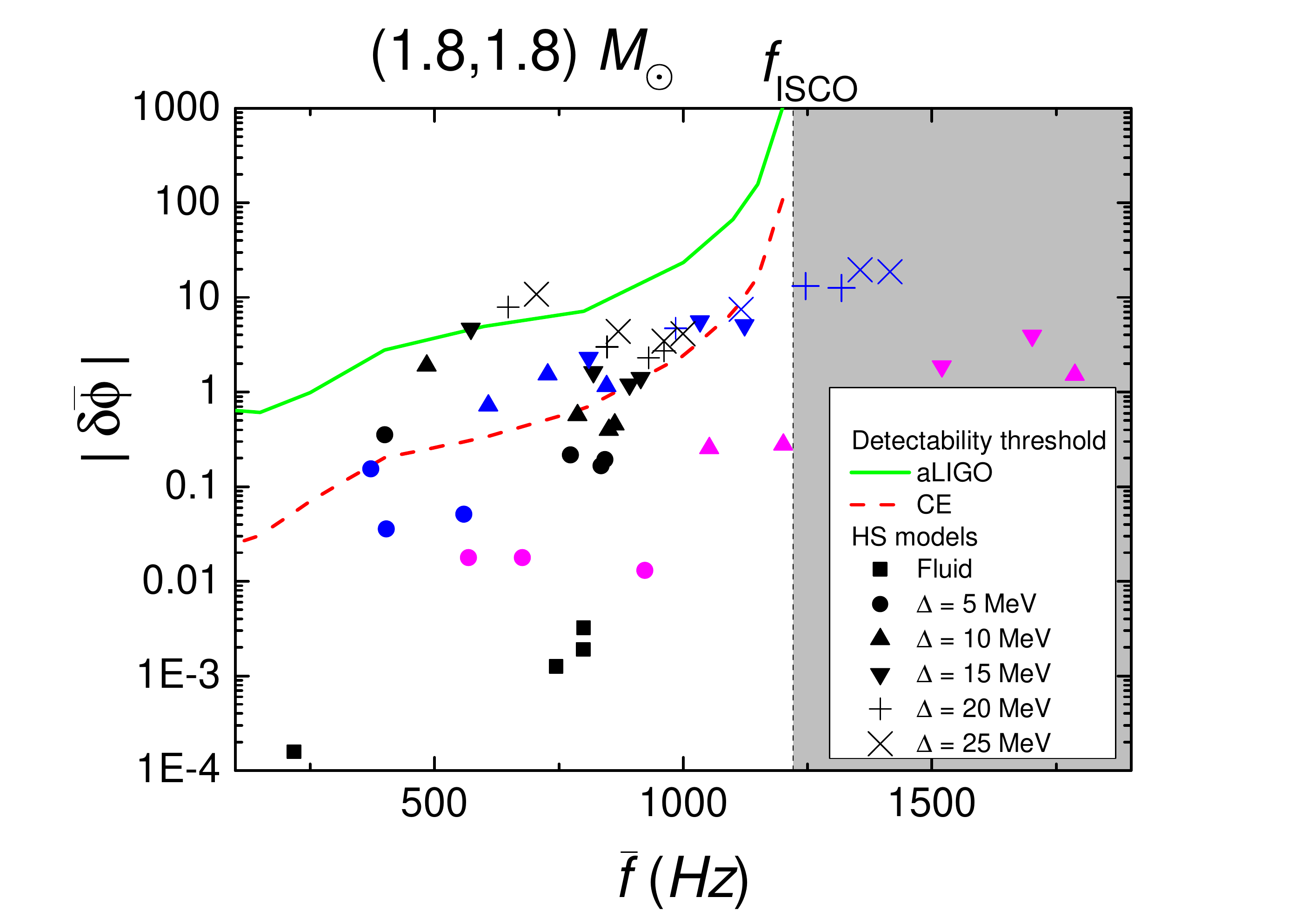}   \caption{\label{Figure_HS_Detectability_14} 
 (Left) The magnitude of the $i$-mode's total overall phase shift $|\delta\bar{\phi}|$ and the corresponding weight-averaged resonant frequency $\bar{f}$ for each HS EOS from Table~\ref{Table_1}, together with the detectability threshold with aLIGO (green solid) and CE (red dashed). If a point is above these curves, such an effect is detectable with the corresponding detector. Here we have assumed an equal-mass HS system with an individual mass of 1.4~$M_\odot$. We consider intermediate $P_t$ models (Heb3-QM-3, Heb3-QM-2, DD2-QM in blue) and low $P_t$ models (MS1-QM, Heb3-QM-1, NL3-QM, TM1-QM in black). 
   The $i$-mode becomes undetectable if the frequency $\bar f$ is higher than the inspiral cutoff frequency (shaded region) that we choose to be at ISCO ($f_{\text{ISCO}})$.
    (Right) Similar to Fig.~\ref{Figure_HS_Detectability_14} but for individual masses of 1.8~$M_\odot$. We also present the high $P_t$ models (MPa1-QM, Heb2-QM, DDH$\delta$-QM) in magenta symbols. The SNR for the $1.4~M_\odot$ system is 440 for aLIGO and 620 for CE, and that of the $1.8~M_\odot$ system is 17 for aLIGO and 760 for CE respectively.
 }
\end{figure*}

In the left panel of Fig.~\ref{Figure_HS_Detectability_14}, we show $|\delta \bar{\phi}|$ and the corresponding $\bar{f}$ for each of the HS models from Table~\ref{Table_1} with mass fixed at 1.4~$M_\odot$, together with the minimum $|\delta \bar{\phi}|$ required for detection based on the Fisher analysis using the Advanced LIGO (aLIGO) \cite{2015CQGra..32g4001L} with its design sensitivity and the Cosmic Explorer (CE) \cite{Abbott_2017}. As discussed above, we set the minimum required $|\delta \bar{\phi}|$ to be its root-mean-square error $\Delta(\delta \bar{\phi})$, obtained from Eq.~\eqref{Fisher_uncertainty}. We have assumed the luminosity distance, $D_L$, to be 100~Mpc. This corresponds to a signal-to-noise (SNR) ratio of 440 for aLIGO and 620 for CE. To account for the number of interferometers, we set $N=2$ for aLIGO and $N=1$ for CE\footnote{The amplitude of GWs is effectively enhanced by $\sqrt{N}$.}. The detection threshold for $|\delta \bar{\phi}|$ increases with $\bar{f}$ because the detector sensitivity deteriorates at higher $f$ and the $i$-mode contributes to the phase only for $f \geq \bar f$ (see Eq.~\eqref{phi_resonance2}) and thus its contribution becomes smaller for higher $\bar f$.

Based on our results, the $i$-mode of some of the low $P_t$ models with large $\Delta$ causes a large $|\delta \bar{\phi}|$ ($\sim10$) in the waveform, making its phase shift above the minimal threshold required for detection with aLIGO. Models with lower $\Delta$ are still above the detectability threshold of CE except for those with zero or very small $\Delta$. As for the intermediate $P_t$ EOSs, the $i$-mode of all the models cannot be detected with the aLIGO detector. With CE, the $i$-mode of a few models within a narrow range of $\Delta$ are detectable. The cutoff frequency, $f_\text{ISCO}$ (see Sec.~\ref{Sec:Fisher}), is also indicated in the figure with a vertical dashed line. The $i$-modes with resonant frequency above this limit cannot be detected from the inspiral signal alone. Since the $i$-mode frequency of the intermediate $P_t$ models depends strongly on $\Delta$ as illustrated in Fig.~\ref{Figure_f_vs_gap}, models with $\Delta$ larger than 15~MeV are beyond this cutoff frequency. Hence, only a few models with $\Delta$ between 5 to 15~MeV have the $i$-mode detectable with CE. For high $P_t$ EOSs, since the central pressure is below $P_t$ for models with 1.4~$M_\odot$, there is no $i$-mode being excited and therefore are not present in the figure. 

We also consider the HS binaries consisting of two 1.8~$M_\odot$ HSs with the results shown in the right panel of Fig.~\ref{Figure_HS_Detectability_14}. Compared to the 1.4~$M_\odot$ case, the low $P_t$ models have lower $|\delta \bar{\phi}|$ in general, while that of the intermediate $P_t$ models are within the same order of magnitude. 
Most of the HS models are below the detectability threshold of the aLIGO detectors except for a few low $P_t$ models with large $\Delta$, while there is still a considerable portion of the low $P_t$ and intermediate $P_t$ models within the detectable region of CE.

The 1.8~$M_\odot$ models with high $P_t$ EOSs have a phase transition at the core, unlike the 1.4~$M_\odot$ models. These models, represented by magenta symbols in the right panel of Fig.~\ref{Figure_HS_Detectability_14}, have low $|\delta \bar{\phi}|$ and are below the detectability threshold of both detectors. The points of the intermediate $P_t$ models are less scattered than the 1.4~$M_\odot$ case, indicating a weaker dependence of $\bar{f}$ and $|\delta \bar{\phi}|$ on $\Delta$. In contrast, the high $P_t$ models show a widespread along $\bar{f}$, which is similar to the case with the 1.4~$M_\odot$ intermediate $P_t$ models.

From the above discussion, we see that $\Delta$ affects the detectability in different ways depending on $P_t$. As $\Delta$ increases, the models in Fig.~\ref{Figure_HS_Detectability_14} shift towards larger values of $\bar{f}$ and $|\delta \bar{\phi}|$ in general. For the low $P_t$ models, the $i$-mode frequency is generally below $f_\text{ISCO}$ for the range of $\Delta$ corresponding to the CCS phase. Hence, the large $\Delta$ models would be more detectable due to their larger phase shift magnitude. On the other hand, the $i$-mode frequency of the high $P_t$ models is more sensitive to $\Delta$. Some models with large $\Delta$ have the $i$-mode frequency higher than $f_\text{ISCO}$, which means the mode is not excited during the inspiral stage. As a result, the models with large $\Delta$ have a higher chance of being detected for the low $P_t$ EOSs, while those with intermediate $\Delta$ are the most detectable ones for the intermediate $P_t$ EOSs. This is similar to the case with 1.4~$M_\odot$ HSs.

\begin{figure*}[!ht]
    \centering
    \includegraphics[width=8.6cm]{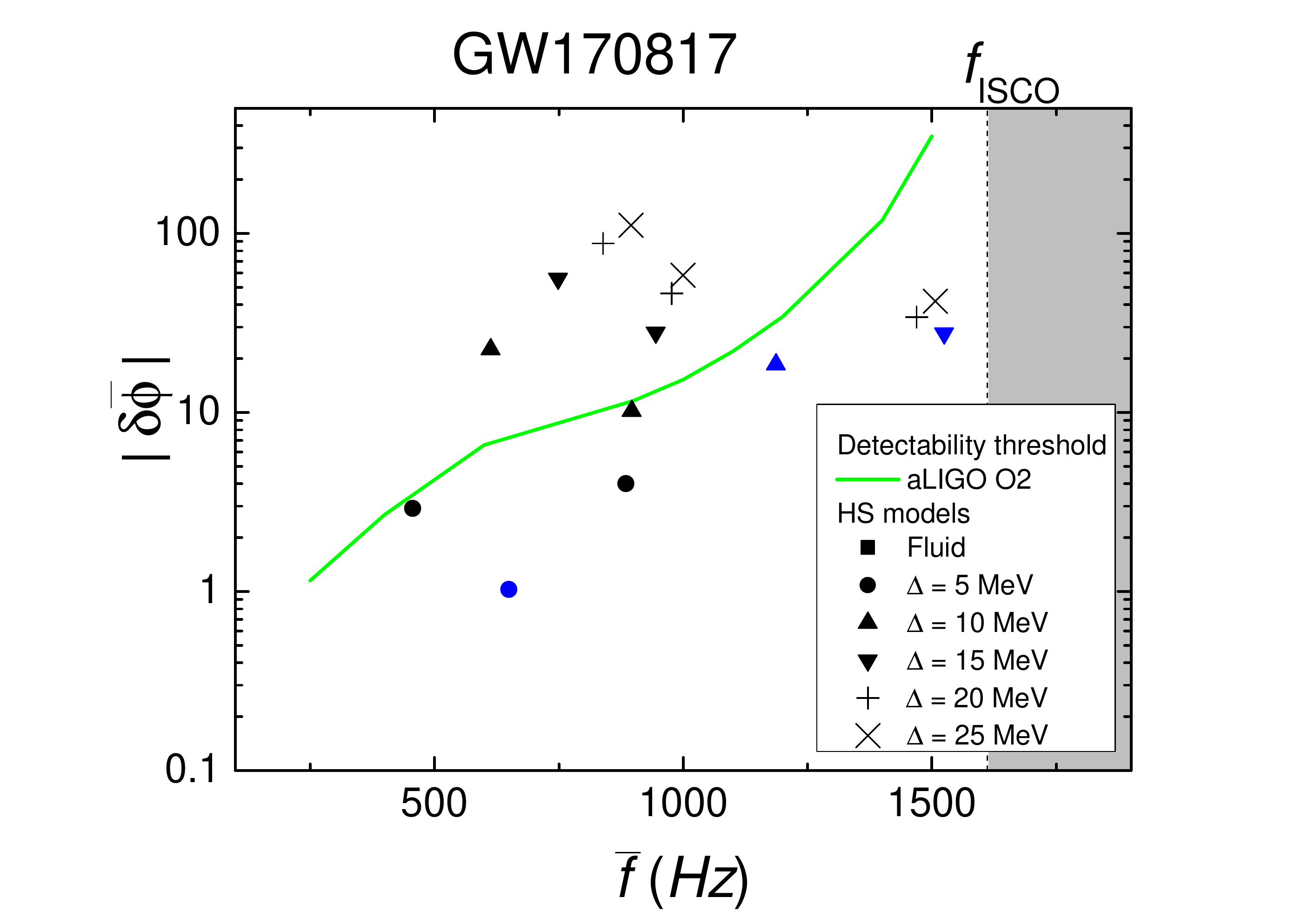}
\includegraphics[width=8.6cm]{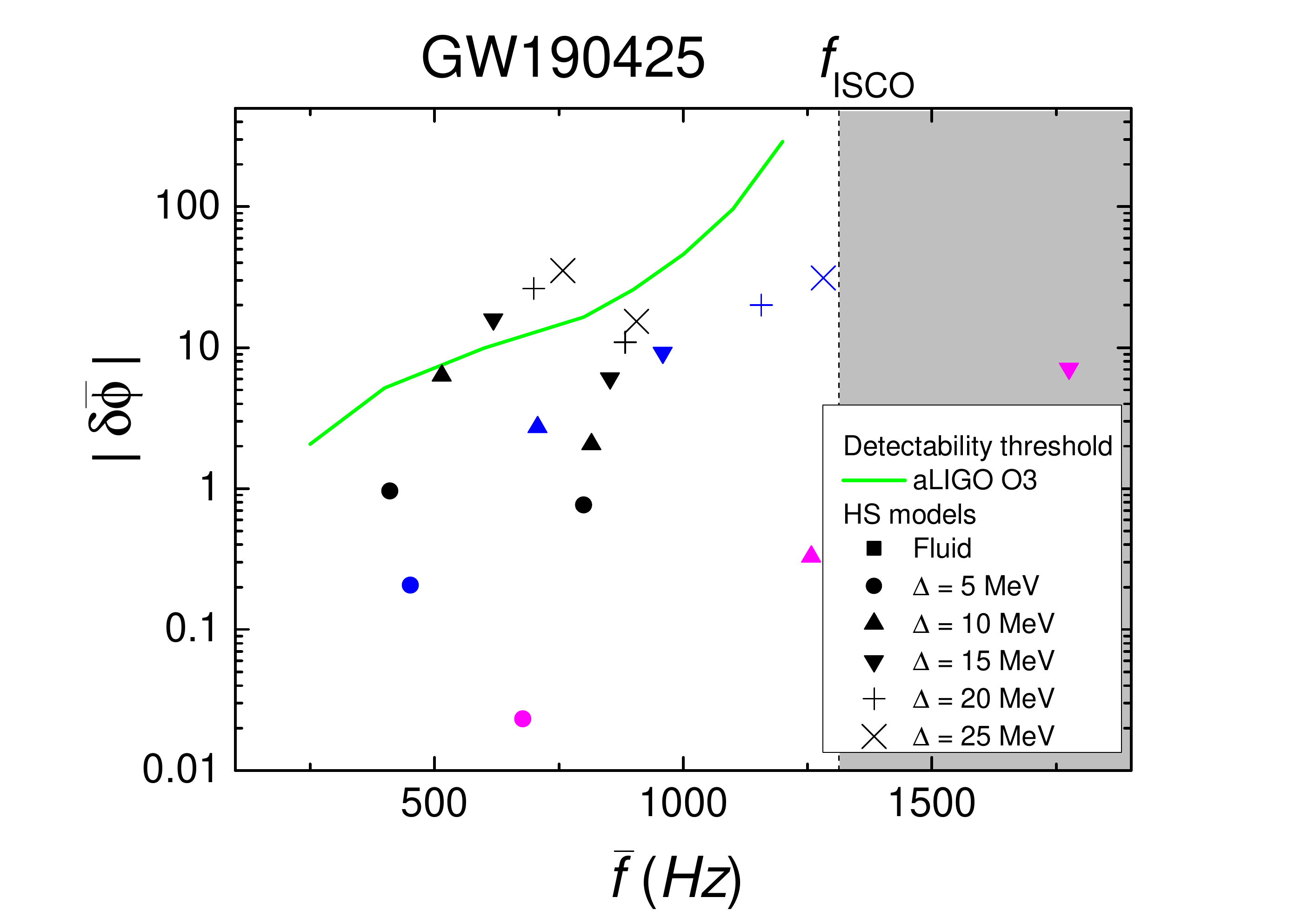}
\caption{\label{Figure_HS_Detectability_O2} 
    (Left) Similar to Fig.~\ref{Figure_HS_Detectability_14} but for parameters consistent with GW170817. The detection threshold curve is computed with the noise curve of aLIGO O2 run. We present the intermediate $P_t$ models (Heb3-QM-3 in blue) and low $P_t$ models (MS1-QM, Heb3-QM-1 in black).
   (Right) Similar to the left panel but for parameters consistent with GW190425. The detection threshold curve is computed with the noise curve of the aLIGO O3 run. We present the high $P_t$ models (MPa1-QM in magenta), the intermediate $P_t$ models (Heb3-QM-3 in blue) and low $P_t$ models (MS1-QM, Heb3-QM-1 in black).
   }
\end{figure*}

\subsubsection{GW170817 and GW190425}

Let us now study the GW events that have been detected, in particular GW170817 and GW190425 that are considered as binary NS mergers. If at least one of the stars in these events has a quark-hadron phase transition, the excitation of the $i$-mode will be encoded in the phase of the inspiral signal. We can apply the method from the previous subsection to analyze its detectability with the corresponding aLIGO run. In our Fisher analysis which gives us the threshold values of $|\delta \bar{\phi}|$, the parameters of the signal ($m_1$, $m_2$, $D_L$) are taken to be (1.46~$M_\odot$, 1.27~$M_\odot$, 40~Mpc) for GW170817 and (1.60~$M_\odot$, 1.75~$M_\odot$, 159~Mpc) for GW190425. The tidal deformability parameters $\bar{\Lambda}$ and $\delta \bar{\Lambda}$ are also adjusted accordingly. We take ($\bar{\Lambda}$, $\delta\bar{\Lambda}$) to be (588, 94) for GW170817 and (160, -20) for GW190425, which are computed with the formulation described in \cite{Hinderer_2008}, assuming the Heb3-QM-1 EOS with a fluid core. Nevertheless, due to their negligible correlations with the $i$-mode parameters, fiducial values of the tidal parameters should not have any significant impact on the numerical results. The noise spectral density data corresponding to the aLIGO second Observing run (O2) for GW170817 and the third Observing run (O3) for GW190425 respectively are obtained from~\cite{ligo_noise_curve}.
We select the high $P_t$ model MPa1-QM, intermediate $P_t$ model Heb3-QM-3 and low $P_t$ models M09m, Heb3-QM-1 from Table~\ref{Table_1} for the analysis. 

The left panel of Fig.~\ref{Figure_HS_Detectability_O2} presents the detectability of HS models for GW170817  with the $i$-mode excitation during the inspiral.
Part of the low $P_t$ models with large $\Delta$ have $|\delta \bar{\phi}|$ above the detectability threshold. Certain models, even having a large $\Delta$, are below the threshold due to the high resonant frequency.
It is worth noting that the values of $|\delta \bar{\phi}|$ can go as high as $\sim$100 for large $\Delta$, which is comparable to that of the $f$-mode (see, e.g., \cite{10.1093/mnras/270.3.611,10.1093/mnras/275.2.301}, for values of $|Q_{nlm}|$). 
The large value of the phase shift mainly comes from the secondary (1.27~$M_\odot$) HS in the binary. The intermediate $P_t$ models have a smaller $i$-mode phase shift in general, and are below the threshold. The strong $\Delta$ dependence of the $i$-mode frequency for the intermediate $P_t$ models makes the frequency go beyond $f_\text{ISCO}$ when $\Delta$ is larger than 15~MeV. 

Meanwhile, the models with the high $P_t$ EOSs do not excite an $i$-mode during inspiral as they consist only of hadronic matter. These findings mean that if GW170817 consists of HSs with a low $P_t$ EOS, it might be possible detect such a feature by performing a data analysis on the GW170817 data similar to that in~\cite{Weinberg:2018icl,Essick:2018wvj,Reyes:2018bee,Pan:2020tht}, given that the CCS $\Delta$ has a value larger than 10~MeV. 
On the other hand, if such an effect is absent, one should be able to constrain the parameter space of $P_t$ and $\Delta$ of the HS EOSs provided we have reasonably good knowledge on the other EOS parameters.

In comparison, the right panel of Fig.~\ref{Figure_HS_Detectability_O2} presents the results for GW190425. Observe that the detection threshold curve for this case is higher due to the increased luminosity distance (and smaller SNR). 
The low $P_t$ models also have smaller $|\delta \bar{\phi}|$, and some of the models are only marginally above the threshold. 
The intermediate $P_t$ models and high $P_t$ models are both below the threshold curve. This agrees with our finding in the previous subsection, that $|\delta \bar{\phi}|$ decreases for larger $P_t$ in general. Besides, the secondary star with 1.60~$M_\odot$ with the high $P_t$ EOS have a central pressure lower than $P_t$ and therefore does not have a quark matter core. Therefore, only the primary star with a higher mass (1.75~$M_\odot$) in the binary contributes to the $i$-mode phase shift. This further lowers the value of $|\delta \bar{\phi}|$ of the high $P_t$ models. Moreover, the higher mass of HSs in the binary leads to a smaller $f_{\text{ISCO}}$, which makes the detection of the $i$-mode challenging for high resonant frequencies.

\section{\label{Sec:ConsistencyCheck} Consistency check of the hybrid method}

In this section, we comment on the validity of the hybrid method that we employed in our analysis.
We solved the TOV equation to construct accurate HS background models, while we used \emph{Newtonian} pulsation theory to compute the $i$-modes for simplicity and applied the method in \cite{10.1093/mnras/270.3.611} to compute the tidal coupling. Ideally, one should compare the results from such an approximate, hybrid method against a fully consistent analysis that solves relativistic perturbation equations. However, given that the framework for solving the latter has not been established yet, we instead follow Yu \textit{et~al.} \cite{10.1093/mnras/stx1188} and compare the hybrid method against a fully-Newtonian one in which both the background and perturbation equations are solved within Newtonian gravity\footnote{We should also emphasize that the full Newtonian approach, despite being consistent throughout the background, pulsation modes and tidal coupling calculations, is not the so-called ``consistent" approach either since the background structure is not accurately determined. The Newtonian treatment in the pulsation mode and tidal coupling problem is also expected to have discrepancies of the size of $M/R$ compared to that of the fully-GR formalism. }. Such a study allows us to estimate the  relativistic effect (in the background solution).

To be more precise, Yu \textit{et~al.} \cite{10.1093/mnras/stx1188} studied the detectability of dynamical tides for hyperon stars.
They compared the deviation in the $g$-mode tidal coupling coefficient $Q_{nlm}$ calculated with the hybrid method from that calculated with a full Newtonian approach and found that $Q_{nlm}$ was off by less than $5\%$. Since the normalization of the eigenmodes in their study contains the mode frequencies,  $Q_{nlm}$  also has a different normalization constant compared to our definition (see Eq.~\eqref{eq:Tidal_Coef}). We therefore compare the estimate of $\delta \phi$ in the two methods, which is independent of the normalization. 

Table~\ref{Table_4} compares the oscillation properties ($f$, $|Q_{n_i22}|$ and $|\delta \phi|$) computed with the hybrid and Newtonian methods. We fix the stellar mass at $1.4M_\odot$ and use the EOS Heb3-QM-1\footnote{We choose Heb3-QM-1 out of the four low $P_t$ EOS due to its lower $i$-mode frequency. For the other EOSs, there are avoided crossings as we change $\Delta$ between the $i$-mode and other modes which distort the wavefunction.} while varying $\Delta$. Notice that the phase shift magnitude computed with the hybrid approach is smaller than that of the full Newtonian approach by a factor of a few. On the other hand, the difference in the oscillation frequency between the two methods is about $25$\%. Although the discrepancy in $\delta \phi$ between the two approaches for the $i$-mode is larger than that for the $g$-modes in \cite{10.1093/mnras/stx1188}, it is still within the same order of magnitude. 
Meanwhile, the $i$-mode phase shift changes by orders of magnitude as we vary the EOSs. Therefore, we expect that the discrepancy does not significantly affect our conclusion except for the marginal cases and we consider the hybrid approach to be a valid order of magnitude estimate of the phase shift. We leave the consistent analysis in full GR for future work.

\begin{table}[!htbp]
\begin{ruledtabular}
	\begin{tabular}{c c c c c c}
	    $\Delta$ (MeV) & Method %& R (km)
	    & $f (\text{Hz}) $& $|Q_{n_i 2 2}|$ & $|\delta \phi|$ \\ 
		\hline 
		5 & Full Newtonian  & 584.37 & 0.040 & 3.167  \\ 
		& Hybrid & 443.03 & 0.020 & 1.136\\ \hline
		15 & Full Newtonian & 1020.3 & 0.295 & 55.853  \\ 
		& Hybrid & 714.86 & 0.143 & 21.901\\ \hline
		25 & Full Newtonian & 1128.4 & 0.399 & 83.460  \\ 
		& Hybrid & 863.44 & 0.248 & 45.009 \\
	\end{tabular} 
\end{ruledtabular}
    \caption{\label{Table_4} The comparison of the numerical results of the 1.4~$M_\odot$ models with the EOS Heb3-QM-1 with a full Newtonian calculation and hybrid approach (TOV equations for background and Newtonian equations for pulsation and tidal coupling). Notice that the frequencies differ by about 25~\% and phase shifts are off by a factor of a few.}
\end{table}

\section{Conclusion}
\label{sec:conclusion}

In this paper, we considered the $i$-mode of HSs with a CCS quark matter core and a hadronic matter envelope, which features an extremely rigid solid core and a fluid envelope. The phase transition is assumed to be first order with a density discontinuity. We studied the resonant excitation of the $i$-mode in HS-HS binary mergers during the inspiral and the corresponding phase shift on the emitted GW waveforms. We then estimated its detectability using a Fisher analysis.

We found that the $i$-mode resonant frequency and the phase shift are rather sensitive to change in the shear modulus of the CCS phase as well as $P_t$, the pressure corresponding to the first-order phase transition. We also found that the chance of detecting the $i$-mode is higher for EOSs with low $P_t$. 
For such low $P_t$ models, the phase shift of the $i$-mode can be above the detection threshold limit if $\Delta$ is large enough, even for GW170817  (Fig.~\ref{Figure_HS_Detectability_O2}).

For the intermediate $P_t$ EOSs, we showed that the sensitivity of aLIGO was insufficient to detect the $i$-mode due to the smaller magnitude of the phase shift. With the third-generation detectors like the CE, a portion of the models with intermediate values of $\Delta$ can be detected. However, those with a large $\Delta$ have high $i$-mode frequencies above the cutoff frequency for inspiral phase and therefore the mode is not excited.
For the high $P_t$ EOSs, quark matter appears inside the core only when the model has a high central pressure, namely the low mass models are simply hadronic NSs without a quark-hadron transition. Focusing on the high mass HSs with a high $P_t$ phase transition, we found that the $i$-mode phase shift of such models is below the detectability threshold of the CE.

Lastly, we comment on the validity of the method we used to study the $i$-mode excitation. We have applied a hybrid method in calculating the $i$-mode excitation by tidal coupling in the binary system by combining a GR background model with Newtonian pulsation equations and Newtonian tidal coupling equations. We estimated the impact of relativistic effects by comparing the results for our hybrid method against the ones from a fully-Newtonian framework. We found that the $i$-mode phase shift can be underestimated by a factor of a few with our method compared to a full Newtonian approach. Therefore, the results presented here should be valid as order-of-magnitude estimates and should not severely affect our conclusion. We note that the full Newtonian method might be less accurate in determining the tidal coupling coefficient than the hybrid one due to the discrepancy between the Newtonian background and the GR background. After all, a fully consistent GR method is required to accurately determine the detectability of the $i$-mode in HSs, which we leave for future work.

There are other avenues for improving the current study in the future. For example, it might be interesting to analyze the actual GW data from binary NS merger events including the $i$-mode contribution in the waveform to place constraints on the quark-hadron phase transition. 
We should also perform a Bayesian analysis for more accurate analysis as the Fisher method adopted here is only valid for events with high SNRs. Furthermore, we may need to relax the adiabatic approximation used for the resonant waveform in this paper, as such an approximation may become invalid if the resonant frequency is too high and the resonant width becomes too large.

We also expect that similar analysis of the $i$-mode can be performed on HS models with a ``Gibbs"-like phase transition instead of the ''Maxwell"-like transition studied here. In such a model, instead of a sharp change in density, a mixed phase is present between the hadronic phase and the quark matter phase. The density is continuous between the interfaces. This phase is highly inhomogeneous and exists as a rigid solid layer (\cite{PhysRevD.46.1274,SOTANI201337}). As a result, there would be two solid-fluid interfaces between the mixed phase and the two pure phases (the quark matter and hadronic matter phases respectively). This is expected to give rise to two $i$-modes, each corresponding to one interface, for each spherical degree $l$. If the $i$-modes are detectable in the GW waveform, this might help further resolve the ``masquerade problem" of the HSs, especially those with mixed phase \cite{2007Natur.445E...7A,Alford_2005}, which states that the macroscopic parameters (mass, radius, tidal deformability, etc) of a HS may be indistinguishable from a hadronic NS. 

\acknowledgments
We thank Lap-Ming Lin, Phil Arras, B\'eatrice Bonga, Zhen Pan, and Huan Yang for carefully reading the manuscript and giving us valuable feedback.
S.Y.L and K.Y. acknowledge support from NASA Grant 80NSSC20K0523. K.Y. further acknowledges support from NSF Award PHY-1806776,  a Sloan Foundation Research Fellowship and the Owens Family Foundation.
K.Y. would like to also acknowledge support by the COST Action GWverse CA16104 and JSPS KAKENHI Grants No. JP17H06358.

\appendix

\section{\label{Newt_puls} Newtonian pulsation equations}

The equations of motion governing the motion of a mass element of an elastic solid consist of the momentum conservation equation, continuity equation and the Poisson equation:
\begin{align}
    \rho \frac{\partial \Vec{v}}{ \partial t} &= \nabla \cdot \bm{S} - \rho \nabla \Phi, \label{EOM_01} \\
    \frac{\partial \rho }{\partial t} &= -\nabla \cdot \left(\rho \Vec{v}\right), \label{EOM_02} \\
    \nabla^2 \Phi &= 4 \pi \rho . \label{EOM_03}
\end{align}
Here $\Vec{v}$ is the velocity vector of the mass element, $\Phi$ is the gravitational potential while $\bm{S}$ is the stress tensor. For an isotropic medium, it is given by
\begin{align}
    S_{ij} = \Gamma_1 P  \, \text{Tr}\left( \mathbf{\epsilon} \right)\delta_{ij}+ 2 \mu \left[ \mathbf{\epsilon}_{ij} - \frac{1}{3} \text{Tr} \left( \mathbf{\epsilon} \right) \delta_{ij} \right].
\end{align}
$\epsilon_{ij}$ is the symmetric strain tensor, $\Gamma_1$ is the adiabatic index defined by $\Gamma_1 =\frac{\rho}{P} \left(\frac{\partial P}{\partial \rho}\right)$ for a fixed entropy and $\mu$ is the shear modulus of the isotropic elastic medium. The fluid limit can be obtained by setting $\mu \to 0$ in the equation of motion.

Next, we decompose the displacement vector and perturbed scalar quantities as follows. The displacement vector of a mass element under spheroidal oscillation is given by 
\begin{align}
    \Vec{\xi} &= \sum_{l, m} \left[ \xi_r^{l}(r) \hat{r} + r \xi_\perp^{l}(r)  \nabla \right] Y_{lm}(\theta, \phi) e^{-i \omega t},
\end{align}
where $\xi_r^{l}(r)$ and $\xi_\perp^{l}(r)$ are the radial  and tangential displacement functions of degree $l$ respectively, while $Y_{lm}(\theta, \phi)$ are the spherical harmonics and $\omega$ is the angular frequency. The Eulerian perturbation of the scalar quantities $\rho$ and $P$ are also expanded in terms of the spherical harmonics:
\begin{align}
    \delta \rho (r, \theta, \phi) &= \sum_{l, m} \delta \rho_{l}(r) Y_{lm}(\theta, \phi), \\
    \delta P (r, \theta, \phi) &= \sum_{l, m} \delta P_{l}(r) Y_{lm}(\theta, \phi), \\
    \delta \Phi (r, \theta, \phi) &= \sum_{l, m} \delta \Phi_{l}(r) Y_{lm}(\theta, \phi).
\end{align}
From now on, we will suppress the spherical degree $l$ in the radial components of the perturbed quantities.

Substituting the perturbed quantities into Eqs.~(\ref{EOM_01})-(\ref{EOM_03}) and using $\vec v = \dot{\vec \xi}$, we obtain the equation of motion of spheroidal pulsation modes. For numerical computation, the radial part of the equations are cast into a system of six coupled ordinary differential equations:
\allowdisplaybreaks
\begin{widetext}
\begin{eqnarray}
r \frac{d z_1}{d r}=&&-\left(1+2\frac{\alpha_2}{\alpha_3}\right)z_1 +\frac{1}{\alpha_3}z_2 +l\left(l+1\right)\frac{ \alpha_2}{\alpha_3}z_3, \label{Pul_01}\\
r \frac{d z_2}{d r} =&&\left(-c_1 V \Omega^2-4V+UV+12\Gamma_1 \frac{\alpha_1}{\alpha_3}\right)z_1 +\left(V - 4 \frac{\alpha_1}{\alpha_3}\right)z_2+l\left(l+1\right) \left( V - 6 \Gamma_1 \frac{\alpha_1}{\alpha_3}\right)z_3 +l\left(l+1\right) z_4 + V z_6, \nonumber \label{Pul_02}\\
\\
r \frac{d z_3}{d r} =&& -z_1 + \frac{1}{\alpha_1} z_4, \label{Pul_03}\\
r \frac{d z_4}{d r} =&& \left(V - 6 \Gamma_1 \frac{\alpha_1}{\alpha_3}\right)z_1 -\frac{\alpha_2}{\alpha_3} z_2+ \left\{-c_1 V \Omega^2  +\frac{2}{\alpha_3}\left[\left(2 l(l+1) -1\right)\alpha_1 \alpha_2 + 2\left(l(l+1) - 1\right)\alpha_1^2\right]\right\}z_3 + \left(V - 3\right) z_4 + V z_5, \label{Pul_04}\\
r \frac{d z_5}{d r} =&& \left(1 - U\right) z_5 + z_6, \label{Pul_05}\\
r \frac{d z_6}{d r} =&& U \left(-A r + \frac{V}{\Gamma_1} - 2 + 2 \frac{\alpha_2}{\alpha_3}\right) z_1 - \frac{U}{\alpha_3}z_2  +l\left(l+1\right) U \left(1 - \frac{\alpha_2}{\alpha_3}\right) z_3 + l \left(l+1\right) z_5  - U z_6, \label{Pul_06}
\end{eqnarray}
\end{widetext}
where the dependent variables $z_1$ to $z_6$ are defined as 
\allowdisplaybreaks
\begin{align}
    z_1 &= \frac{\xi_r}{r}, \\
    z_2 &= \alpha_2 \left[\frac{ 1}{r^2}\frac{d}{dr}\left( r^2 \xi_r\right) -\frac{l(l+1)}{r}\xi_\perp\right] + 2 \alpha_1 \frac{d \xi_r}{d r},\label{def_z2}\\
    z_3 &= \frac{\xi_\perp}{r}, \\
    z_4 &= \alpha_1 \left( \frac{d \xi_\perp}{dr} - \frac{\xi_\perp}{r} + \frac{\xi_r}{r}\right),\\
    z_5 &= \frac{\delta \Phi}{g r}, \\
    z_6 &= \frac{1}{g}\frac{d \delta \Phi}{dr},
\end{align}
$g$ is the Newtonian gravitational acceleration given by $m/r^2$ and the functions $\Omega$, $c_1$, $\alpha_1$, $\alpha_2$, $\alpha_3$, $A$, $U$ and $V$ are defined as 
\allowdisplaybreaks
\begin{align}
    \Omega &= \sqrt{\frac{R^3 \omega^2 }{M}}, \\
    c_1 &= \left(\frac{r}{R}\right)^3 \frac{M}{m}, \\
    \alpha_1 &= \frac{\mu}{P},\\
    \alpha_2 &= \Gamma_1-\frac{2}{3}\frac{\mu}{P},\\
    \alpha_3 &= \Gamma_1+\frac{1}{3}\frac{\mu}{P},\\
    A &= \frac{1}{\rho} \frac{d \rho}{dr} - \frac{1}{\Gamma_1 P} \frac{d P}{dr},\\
    U &= \frac{r}{m} \frac{d m}{dr}, \\
    V &= -\frac{r}{P} \frac{d P}{dr}.
\end{align}
Here $M$ and $R$ are the stellar mass and radius. $A$ is the Schwarzschild discriminant and it vanishes in cold compact objects except at the density discontinuities. Equations~\eqref{Pul_01}--\eqref{Pul_06} describe the linear perturbations of the HS solid core.
Notice that there are no independent equations for $\delta \rho$ and $\delta P$ since the variable $\delta \rho$ is related to $z_2$ through Eq.~\eqref{def_z2} and the perturbed continuity equation derived from Eq.~\eqref{EOM_02}:
\begin{align}
    \frac{ 1}{r^2}\frac{d}{dr}\left( r^2 \xi_r\right) -\frac{l(l+1)}{r}\xi_\perp = -\frac{\Delta \rho}{\rho},
\end{align}
whereas the variable $\delta P$ can be related to $\delta \rho$ through the linearized thermodynamic identity: 
\begin{align}
\frac{1}{\rho}\Delta \rho = \frac{1}{ \Gamma_1 P}\Delta P.
\end{align}
Here $\Delta f$ represents the Lagrangian perturbation of a variable $f$ depending on $r$, which is related to the Eulerian perturbation $\delta f$ by
\begin{align}
    \Delta f = \delta f + \xi_r \frac{df}{dr}.
\end{align}

To numerically obtain the pulsation modes, we integrate Eqs.~(\ref{Pul_01})--(\ref{Pul_06}) from the center to the stellar radius. At the solid-fluid interface, we employ continuity conditions of the pulsation variables $z_1$, $z_2$, $z_4$ and $z_5$. The continuity of $z_1$ is the direct consequence of the assumption that the volume element at the interface contains no void if the phase transition happens slowly compared to the pulsation motion (see e.g., \cite{Pereira_2018}). The continuity of $z_2$ and $z_4$ comes from the continuity of the stress in the radial and tangential directions. Lastly, the Poisson equation guarantees the continuity of $z_5$. \footnote{Note that $z_3$ and $z_6$, which are related to the tangential displacement and the first derivative of the gravitational potential perturbation respectively, are not required to be continuous. The former is the consequence of the so-called ``free-slipping" condition and the latter is allowed by the Poisson equation.}

The above equations for $z_1$--$z_6$ describe the pulsation problem of the solid core. Although we can  in principle obtain the pulsation equations inside fluid by taking the $\mu \rightarrow 0$ limit, it is straightforward to see that Eqs.~\eqref{Pul_03} and \eqref{Pul_04} become trivial in this limit and we effectively have only four coupled differential equations. Therefore, it is often better to introduce another set of dependent variables for the fluid problem. Inside the fluid envelope, we employ the formulation by \cite{1971AcA....21..289D} (see also P.225 of \cite{1980tsp..book.....C}):
\begin{align}
r \frac{d y_1}{d r} &= \left(\frac{V}{\Gamma_1}-3\right)y_1 +\left[\frac{l \left(l+1\right)}{c_1 \Omega^2} - \frac{V}{\Gamma_1}\right]y_2 + \frac{V}{\Gamma_1}y_3,\label{pul_y1}\\
r \frac{d y_2}{d r} &=  \left(c_1 \Omega^2 + A r\right)y_1 + \left(1 - U - A r\right)y_2 + A r y_3,\label{pul_y2}\\
r \frac{d y_3}{d r} &= \left(1 - U\right) y_3 + y_4,\label{pul_y3}\\
r \frac{d y_4}{d r} &= - U A r y_1 + \frac{U V}{\Gamma_1} y_2 + \left[l \left(l+1\right) - \frac{U V}{\Gamma_1}\right]y_3 - U y_4. \label{pul_y4}
\end{align}
Here the pulsation variables are given by 
\begin{align}
y_1 &= z_1 = \frac{\xi_r}{r}, \\
y_2 &= \frac{1}{g r}\left(\frac{\delta P}{\rho} + \delta \phi\right), \label{PulDef_y2}\\
y_3 &= z_5 = \frac{\delta \Phi}{g r},\\
y_4 &= z_6 = \frac{1}{g}\frac{d \delta \Phi}{dr}.
\end{align}
$y_2$ is also related to $\xi_\perp$ through
\begin{align}
    y_2 = c_1 \Omega^2 z_3= \frac{\omega^2}{g}\xi_\perp.
\end{align}
Equation~(\ref{PulDef_y2}) implies that the continuity of radial stress across the interface is equivalent to
\begin{align}
    \left[V\left(y_1 - y_2 + y_3\right)\right]_{\text{fluid}}=\left[z_2\right]_{\text{solid}}.\label{eq:interface_y2}
\end{align}
Here the square brackets ``[ ]" with the subscripts ``fluid" or ``solid" indicates that the expression enclosed is evaluated at the fluid side or the solid side of the interface respectively.

To determine $y_4$ at the interface, one last continuity condition is derived by integrating Eq.~\eqref{pul_y4} across the interface, using the fact that the derivative of $\rho$ in $A$ behaves like a Dirac delta function in $r$. Doing so, one can find:
\begin{align}
    \left[U y_1 + y_4\right]_{\text{fluid}} = \left[U z_1 + z_6\right]_{\text{solid}}. \label{eq:interface_y4}
\end{align}
This equation corresponds to the continuity of the Newtonian gravitational force at the perturbed interface.

At the surface, we have similar continuity conditions as Eqs.~\eqref{eq:interface_y2} and \eqref{eq:interface_y4}:
\begin{align}
    y_1 - y_2 + y_3 =& 0, \\
    U y_1 + y_4 =& -\left(l+1\right) y_3,
\end{align}
where all quantities are evaluated at $r=R$. The second equation comes from the continuity of $y_3$ and we have applied the solution to the Poisson equation in vacuum (i.e., $\delta \Phi \propto r^{-l-1} $).

While integrating Eqs.~\eqref{Pul_01}-\eqref{Pul_06} from $r=0$ numerically for the solid core, we consider only the regular solutions, which can be obtained from a Taylor series expansion of $z_1$--$z_6$ near $r=0$. We modify the expressions of the regular solutions derived by \cite{10.1111/j.1365-246X.1975.tb04145.x} to fit our definition of pulsation variables:
\begin{align}
    z_1 =& A_0 r^{l-2} + A_2 r^{l},\\
    z_2 =& B_0 r^{l-2} + B_2 r^{l},\\
    z_3 =& C_0 r^{l-2} + C_2 r^{l},\\
    z_4 =& D_0 r^{l-2} + D_2 r^{l},\\
    z_5 =& \frac{E_0}{g r} r^{l-2} + \frac{E_2}{g r} r^{l},\\
    z_6 =& \frac{1}{g}\left[F_0+3\sigma A_0-(l+1)E_0\right] r^{l-2} \\
    &+ \frac{1}{g}\left[(l+2)F_2-3\sigma A_2\right] r^{l},
\end{align}
where the coefficients are related by
\begin{align}
    A_0 &= l C_0,\\
    B_0 &= 2(l-1) \alpha_1 A_0,\\
    D_0 &= \frac{2\alpha_1(l-1)}{l} A_0,\\
    E_0 &= 3\sigma C_0 + \frac{1}{l}F_0,\\
    C_2 &= \frac{\beta_2}{\beta_1} D_2 + \frac{\rho}{P \beta_1}\left\{F_0+\left[\omega^2 + (3-l)\sigma\right]A_0\right\},\\
    A_2 &= -l C_2 + \frac{1}{\alpha_1}D_2,\\
    B_2 &= \gamma_1 C_2 + \gamma_2 D_2,\\
    E_2 &= \frac{3}{2}\sigma(2l-3)\left[(l+3)A_2-l(l+1)C_2\right],\\
    F_2 &= (l+2)E_2 -3\sigma A_2,
\end{align}
and $\sigma$, $\beta_1$,$\beta_2$, $\gamma_1$ and $\gamma_2$ are given by
\allowdisplaybreaks
\begin{align}
    \sigma =& \frac{4\pi}{3}\rho,\\
    \beta_1 =& 2l^2(l+2)\alpha_2  + 2 l (l^2+2l-1)\alpha_1,\\
    \beta_2 =& l(l+5) + l(l+3)\frac{\alpha_2}{\alpha_1},\\
    \gamma_1 =& 2 l (l+2)\alpha_2  + 2 l(l+1)\alpha_1 ,\\
    \gamma_2 =& 2(l+1) + (l+3)\frac{\alpha_2}{\alpha_1}.
\end{align}
By choosing arbitrary values of $C_0$, $D_2$ and $F_0$ (or any 3 of the 12 coefficients), we can obtain three independent regular series solutions about $r=0$ for the pulsation problem in the solid core.

If we consider HS models with a fluid quark matter core (i.e., the $\Delta = 0$ limit for the CCS phase), we also need the regular solutions for Eqs.~\eqref{pul_y1}--\eqref{pul_y4} near $r=0$. Following \cite{1971AcA....21..289D}, the regular solutions satisfy the following equations
\begin{align}
    y_2 =& \frac{c_1 \omega^2}{l}y_1,\\
    y_4 =& l y_3.
\end{align}
Hence, there are two independent regular solutions at the center.

\section{\label{Maxwell_Construction} Maxwell Construction}
The quark-hadron matter phase transition can either be of first or second order depending on the charge screening effect and the surface tension between the phases. They can respectively be constructed through a Maxwell construction or a Gibbs construction which results in a mixed phase \cite{PhysRevD.46.1274}. 

We focus on the Maxwell construction, which gives a first-order phase transition with a sharp density jump at the transition pressure $P_t$ inside the HS. The transition point is determined by the following equations (\cite{PhysRev.172.1325,Bhattacharyya_2010}):
\begin{align}
P_t =& P_1(\mu_B, \mu_e) = P_2(\mu_B, \mu_e), \\
\mu_B =& \mu_{B1} = \mu_{B2},
\end{align}
where $\mu_B, \mu_e$ are the baryon chemical potential and electron chemical potential. The subscripts 1 and 2 of the pressure indicate the hadronic phase and the quark matter phase respectively. Average chemical potential of quarks $\mu_q$ is given by:
\begin{align}
\mu_q = \frac{\mu_u + \mu_d + \mu_s}{3}.
\end{align}
Since three quarks form one baryon, we can relate the chemical potentials by
\begin{align}
3 \mu_q = \mu_B.
\end{align}
For a given NS EOS, $\mu_B$ can be determined with the Euler equation
\begin{align}
\mu_B = \frac{\rho + P}{n_B},
\end{align}
where $\rho$ is the energy density and $n_B$ is the baryon number density given by
\begin{align}
n_B = n_n + n_p,
\end{align}
where $n_n$ and $n_p$ are the number density for neutrons and protons respectively.

\nocite{*}

\bibliography{apssamp}.

\end{document}